\documentstyle[12pt,epsfig]{article}

\newcommand{\be}{\begin{equation}}
\newcommand{\bea}{\begin{eqnarray}}
\newcommand{\ba}{\begin{array}}
\newcommand{\bean}{\begin{eqnarray*}}
\newcommand{\ee}{\end{equation}}
\newcommand{\eea}{\end{eqnarray}}
\newcommand{\ea}{\end{array}}
\newcommand{\eean}{\end{eqnarray*}}

\def \dsl {\partial \kern-.55em{/}}
\def \Dsl {D \kern-.65em{/}}
\def \nonSM {SM \kern-.75em{/}} 
\def \qsl {q \kern-.45em{/}}
\def \slp {p \kern-.45em{/}}
\def \ksl {k \kern-.45em{/}}
\def \lsl {l \kern-.45em{/}}

\catcode`@=11
\newcount\@tempcntc
\def\@citex[#1]#2{\if@filesw\immediate\write\@auxout{\string\citation{#2}}\fi
  \@tempcnta\z@\@tempcntb\m@ne\def\@citea{}\@cite{\@for\@citeb:=#2\do
    {\@ifundefined
       {b@\@citeb}{\@citeo\@tempcntb\m@ne\@citea\def\@citea{,}{\bf ?}\@warning
       {Citation `\@citeb' on page \thepage \space undefined}}%
    {\setbox\z@\hbox{\global\@tempcntc0\csname b@\@citeb\endcsname\relax}%
     \ifnum\@tempcntc=\z@ \@citeo\@tempcntb\m@ne
       \@citea\def\@citea{,}\hbox{\csname b@\@citeb\endcsname}%
     \else
      \advance\@tempcntb\@ne
      \ifnum\@tempcntb=\@tempcntc
      \else\advance\@tempcntb\m@ne\@citeo
      \@tempcnta\@tempcntc\@tempcntb\@tempcntc\fi\fi}}\@citeo}{#1}}
\def\@citeo{\ifnum\@tempcnta>\@tempcntb\else\@citea\def\@citea{,}%
  \ifnum\@tempcnta=\@tempcntb\the\@tempcnta\else
   {\advance\@tempcnta\@ne\ifnum\@tempcnta=\@tempcntb \else \def\@citea{--}\fi
    \advance\@tempcnta\m@ne\the\@tempcnta\@citea\the\@tempcntb}\fi\fi}
\catcode`@=12
 
\begin{document}

\begin{titlepage}

\begin{flushright}
DTP/98/46\\
January 1999
\end{flushright}
 
 \vspace*{3cm}
 
\begin{center}
{\Large\bf Restoring good high energy behaviour \\ in Higgs 
production via W fusion at the LHC}\\[0.4cm]
{\large K. Philippides$^1$} 
{\large and  W.J.  Stirling$^{1,2}$}\\[0.4cm]
{\em $^1\ $Department of Physics, University of Durham, Durham, DH1 3LE, U.K.}
\\
{\em $^2\ $Department of Mathematical Sciences, University of Durham, Durham, 
DH1 3LE,  U.K.}
\end{center}
 
\vskip1.0cm    
\centerline{\bf   ABSTRACT}  
\vskip0.7cm    

\noindent The $W-$fusion scattering process 
$W^+W^-\!\rightarrow ZZ$ for off-shell $W$ bosons 
 is studied,  focusing on the issue of its high-energy behaviour 
 which is known to be anomalous. It is shown that the unitarity violating 
 terms can be isolated and extracted in a well-defined and efficient 
 way using the pinch- technique. 
This restores the good high energy behaviour of the cross section 
 and, in particular, 
  makes possible the identification of the Higgs resonance in the invariant 
 mass distribution $m_{ZZ}$ of the $Z$ pair. The discarded terms, which are 
  proportional to the off-shellness of the $W$ bosons, cancel against similar 
   terms originating from the remaining diagrams for the full physical 
    process $f_1 f_2\rightarrow \overline{f'}_1 \overline{f'}_2 Z Z$. 
This cancellation ensures the gauge invariance of our result, which therefore 
constitutes a meaningful separation between {\it signal} and {\it background} 
when they both contribute coherently. 
   Equipped with this result, we are able to define a resonant approximation 
   for the process $p p \rightarrow Z Z +\mbox{2 jets} + X ,$ 
which circumvents the 
    problem of good high energy behaviour without having to resort to the 
    lengthy calculation of the complete set of diagrams.  
    In this approximation 
    only the $W-$ and $Z-$fusion {\it signal}
graphs are included, i.e. the ones which 
    contain the Higgs resonance. We have verified that the  
    approximate resonant 
 cross section  describes very well the full 
    result not only close to  the resonance but also beyond it.

\end{titlepage}

\setcounter{equation}{0}

\section{Introduction}
The discovery of the Higgs boson is the primary physics goal of
the LHC $pp$ collider. For low Higgs masses the dominant
production mechanism is gluon fusion, $gg \to H$, but for a heavy Higgs
there is also a significant contribution from the $W-$fusion process,
$WW\to H$, where the $W$s are emitted from incoming 
quarks.\footnote{See, for example,
the review in \cite{BOOK}.} Since a heavy (Standard Model) Higgs is most 
easily detected in the $ZZ$ decay channel, the process of interest
is $WW\to H \to ZZ$. However the $s-$channel Higgs resonant diagram
is only one of several diagrams contributing to this scattering process,
the other diagrams containing, for example,  trilinear and quadrilinear
gauge boson vertices. In fact the $W-$fusion process 
 \be
 W(p_1)W (p_2)\rightarrow Z(k_1)Z (k_2)
 \label{Wfusion}
 \ee
 provides a classic illustration  of the subtle gauge cancellations encountered 
 in non-Abelian gauge theories. The role of the Higgs graph is crucial  
 in  obtaining  a cross section which is well behaved at high energies,
 see for example the discussion in \cite{BOOK}. 
   In practice, however,  one  actually has to consider the case
   of  {\it off-shell} 
   $W-$fusion, since $W$s emitted from the incoming quarks have $q^2 < 0$.
    But when the $W$s 
    are {\it off-shell}  the delicate gauge cancellations responsible 
    for the good high energy behaviour of the amplitude are spoiled. 
    Unitarity is badly violated by terms which are proportional to the 
    {\it off-shellness} of the $W$s \cite{StiKle86},
\bea   
W_1&=& p_1^2-M_W^2 \nonumber \\
W_2&=& p_2^2-M_W^2 ~.
\label{Wprop}
\eea
Unitarity is only restored for 
 the physical cross section when   the {\it full} set of diagrams for the 
 $qq \to qq ZZ$ process is taken into account \cite{StiKle86}, \cite{GloBa}, 
i.e. not just the subset containing
 $WW\to ZZ$. This involves a very large   number of additional Feynman graphs. 

Since in the region of $M_{ZZ} \sim M_H$ the {\it on-shell}
$WW\to ZZ$ proces is clearly well approximated by the 
  $s$-channel Higgs resonance graph (at least for $M_H$ not too large so
  that $\Gamma_H \ll M_H$), it is interesting to ask whether the full
  $qq\to qqZZ$ process, including off-shell $WW$ scattering, 
   can be similarly approximated, while retaining good high-energy behaviour. 
In other words, one seeks a  minimal set of diagrams that include 
 the Higgs resonance while at the same time not spoiling 
 the delicate gauge cancellations.
 
\begin{figure}[ht]
\begin{center}
\mbox{\epsfig{figure=
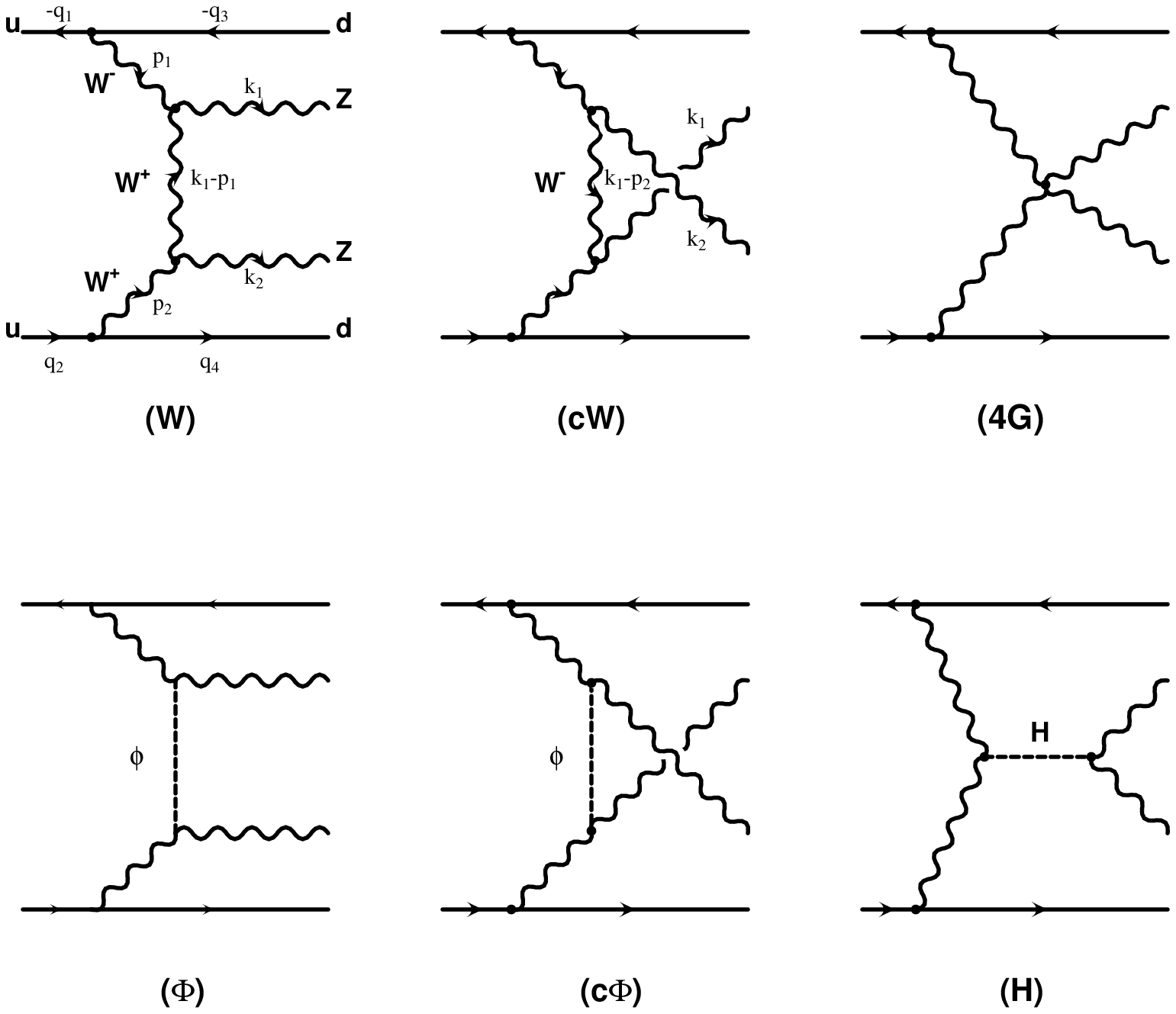,width=15.0cm}}
\label{fig1}
\end{center}
\noindent {\bf Figure 1:} The six $W$ fusion graphs $T_{WW}$ and relevant 
kinematics. 
\end{figure}

To begin with one could  consider only  the 
 Higgs resonant diagram  contributing to $qq\to qqZZ$. 
 However this leads to a
 cross section that 
  behaves badly at high energies, growing  like $s$. As a result 
  the shape of the differential cross section around the resonance 
  cannot be trusted.
    As a next step one could extend the set to contain all six of 
  the $W-$fusion diagrams of process (\ref{Wfusion}).
  As already mentioned,  there {\it are}  potential cancellations 
  at high energy  in this case but the cross section actually 
  respects unitarity only when
    both $W$s are {\it on-shell}. In the {\it off-shell} case
    when the six diagrams are embedded in $qq\to qqZZ$, see Fig.~\ref{fig1},
   this is 
   not true anymore and in fact  
   the situation gets even worse. The cross section now grows as $s^2$ 
   with the leading terms  always proportional to the {\it off-shellness}
   of the $W$s, i.e. $W_1$ and $W_2$.
   For high enough energies the Higgs resonance gets completely swamped 
    by these {\it off-shell} terms and the
    differential cross section exhibits no resonance structure. 
Only when the full set of  graphs (Figs.~1 -- 3)  
  is considered 
 does one obtain a cross section that behaves well at high energies.  
   In this set large gauge cancellations take place 
   and the resulting 
   cross section exhibits the same behaviour as for {\it on-shell} Higgs production
   through the process 
   $qq \rightarrow qq H$,  namely it
    grows slowly  as $\ell n(s)$ for large $s$.

The bad high-energy behaviour of individual diagrams can be attributed
to the following two factors: 
i) the presence of the trilinear gauge boson vertices, and 
ii) the fact that the $Z$ bosons  can be longitudinally polarized. 
Both these factors bring extra momenta in the numerator of Feynman graphs. 
    In fact the polarization 
    vector of an energetic longitudinal $Z$ boson behaves as
    \be
    \epsilon_L^{\mu}(k_1) = \frac{k_1^{\mu}}{M_Z} + {\cal O}(M_Z/E_Z)~.
    \label{eq:zpol}
    \ee
 As a result
of the $M_Z$ factor in the denominator, 
the cross section can grow as powers of $s/M_Z^2$.
 However the momenta introduced by these factors  into the 
  numerator of the  Feynman graphs are at the same time 
 the momenta which  make possible a `communication' between 
 graphs with seemingly different propagator structures. It is this 
 communication that eventually leads to the gauge cancellations between 
 different diagrams that
 restore the good high-energy behaviour. 
      Obviously the cancellation of gauge-dependent terms between different 
   graphs will be even more pronounced in a general $R_{\xi}$ gauge, 
   due to the presence of the extra momenta in the longitudinal part 
   of a gauge boson's propagator in this case.
   
Any attempt  to  approximate Higgs production using only resonant production via
$WW$ fusion (or indeed any other multi-gauge-boson production process) cannot 
therefore succeed unless the above gauge cancellations are correctly taken into account.
The recently developed {\it tree level pinch technique}
 \cite{PRW}, \cite{PiPa},  
provides exactly the right calculational framework for addressing this 
 problem.  It allows a rigorous definition of a gauge invariant sub-amplitude 
 which can be used to approximate a full scattering amplitude.
For example in \cite{PRW},  the authors considered 
the process $e^+e^-\rightarrow W^+W^-$ and were able to 
establish good high-energy
    behaviour for each individual 
    square or interference term of the three contributing Feynman diagrams. 

In this paper we will apply similar methods in order to isolate the $W-$fusion
part of the $q q  \rightarrow q q Z $ process in the form of a squared subamplitude
that is both gauge independent and respects unitarity. This subamplitude squared
can then be used to provide a well-defined approximation to Higgs production
based only on $W-$fusion that, as we shall see, works particularly well in the region
of the resonance and above. It should therefore provide a useful analysis tool
 for simulating Higgs  production at, for example, the LHC $pp$ collider,
 especially in a Monte Carlo context where the compactness of our expressions
 for the subamplitude avoids a time consuming calculation of the full amplitude.
 
 The paper is in essentially two parts. In the following section we discuss
 in some detail how to implement the pinch technique in the context of the 
 $W-$fusion process. Having arrived at the final expressions for the approximate
 scattering amplitude squared, we then perform numerical studies to compare our
 results for various distributions with those obtained from the full calculation.
Our conclusions are presented in a final section.

\setcounter{equation}{0}
\section{
A unitarity respecting amplitude for {\it off-shell} $W$ fusion}
 
In this section we will consider in detail the  process
\be
\bar{u}(q_1)u(q_2) \rightarrow \bar{d}(q_3)d(q_4)Z(k_1)Z(k_2)
\label{fullprocess}
\ee 
which  obviously can proceed via  $W-$ fusion.
We will show how it is possible to isolate the $W-$fusion part of this process
in a well defined way,  and arrive at a squared amplitude that is
gauge independent and respects unitarity. 

The  complete set of Feynman diagrams for this process
exhibits a plethora of different propagators, and we can use these
to classify the various types of diagrams. 
First, we separate all graphs into two categories according to whether the
initial quarks annihilate ($S$ graphs) or not ($T$ graphs).
The $T$ graphs are those that contain $t$-channel $W$ propagators.
 There are in total 34 $T$ graphs (in the Feynman gauge) and we may further
separate them according to their 
fermion propagator structure or, equivalently, according  to the particles
 that emit the final-state  $Z$ bosons.
Thus if we require both $Z$s to be emitted from a gauge boson, we
obtain the usual six $W-$fusion graphs.
These will be denoted collectively by $T_{WW}$
\be
T_{WW}= T_{W}+T_{cW} + T_{4G}+T_{\phi}+T_{c\phi} + T_{H}
\ee
 and are depicted in Fig.~1.
 Their characteristic feature is that they do not contain any 
 fermion propagators. In fact their structure is described by
\be
T_{WW} = \frac{V_{\alpha}}{ W_1}\frac{U_{\beta}}{W_2}
T^{\alpha\beta\mu\nu}_{WW}\epsilon_1^{\mu}\epsilon_2^{\nu} \; ,
\label{WWstruct}
\ee
where $W_1$ and $W_2$ 
are the inverse propagators of the two {\it off-shell} $W$ bosons 
given in Eq.~(\ref{Wprop}).
The external fermions only enter through  the two fermion currents
\bea
V_{\alpha} &=& \frac{g}{\sqrt{2}}
\overline{v}_u(q_1)\gamma_{\alpha}P_L v_d(q_3)\\
U_{\beta}&=& \frac{g}{\sqrt{2}}\overline{u}_d(q_4)\gamma_{\beta}P_L u_u(q_2)
\; .
\eea

In the Feynman gauge the individual graphs are given explicitly by
\bea
T^{\alpha\beta\mu\nu}_W&=& \frac{c_W^2}{D_1}~
\Gamma^{\alpha\rho\mu}(p_1,k_1-p_1,-k_1)
\Gamma^{\beta\rho\nu}(p_2,k_2-p_2,-k_2) \label{W} \\
T^{\alpha\beta\mu\nu}_{cW}&=& \frac{c_W^2}{D_2}~
\Gamma^{\alpha\rho\nu}(p_1,k_2-p_1,-k_2)
\Gamma^{\beta\rho\mu}(p_2,p_2,1-p_2,-k_1) \label{cW}\\
T^{\alpha\beta\mu\nu}_{4G}&=& c_W^2 
G^{\alpha\beta\mu\nu}(p_1,p_2,k_1,k_2)
\label{4Ggraph} \\
T^{\alpha\beta\mu\nu}_{\phi}&=& \frac{M_Z^2s_W^4}{D_1}~
                                  g^{\alpha\mu}g^{\beta\nu}\label{f}\\
T^{\alpha\beta\mu\nu}_{c\phi}&=&\frac{M_Z^2s_W^4}{D_2}~
                                  g^{\alpha\nu}g^{\beta\mu}\label{cf}\\
T^{\alpha\beta\mu\nu}_H &=& \frac{M_Z^2}{D_H}g^{\alpha\beta}g^{\mu\nu}
\label{Higgs}
\eea 
where an overall factor of $(-ig^2)$ has been omitted from all graphs.

 The kinematics of the process is described in Fig.~1, i.e.
\be
p_1 = q_1-q_3\; ,~~~~~~
p_2 = q_2-q_4\;  ,~~~~~~
p_1+p_2 = k_1+k_2=q\; .
\label{kin}
\ee  
The inverse bosonic propagators appearing  in these graphs are
\bea
D_1&=&(k_1-p_1)^2-M_W^2=(k_2-p_2)^2-M_W^2\; ,\\
D_2&=&(k_1-p_2)^2-M_W^2=(k_2-p_1)^2-M_W^2 \; ,\\
D_H&=& q^2-M_H^2\; .
\label{Dprops}
\eea
Note that an imaginary part must be included in the Higgs propagator
$D_H$ in order to regulate it when $q^2=M_H^2$. We will comment on
its precise form later. The  $t$-channel propagators of the $W$s
will always be spacelike ($< 0$). In fact for the two $W$s emitted from the
external fermions we have 
\be
p_1^2=-\frac{\sqrt{s}E_3}{2}(1-\cos\theta_{13})\; ,~~~~~~~~~~~~
p_2^2=-\frac{\sqrt{s}E_4}{2}(1-\cos\theta_{24}) \; ,
\ee
where the energies and scattering angles refer to the centre-of-mass frame.
Thus the $W$s will always be {\it off-shell} by at least an amount equal
to their mass, $\vert W_1\vert, \vert W_2\vert \geq M_W^2$, with the
minimum being attained for forward scattering,
$\theta_{13}=0=\theta_{24}$ ($\Rightarrow\; p_1^2=0=p_2^2$).

 The forms of the trilinear $Z_{\mu}W^-_{\alpha}W^+_{\beta}$ 
and quadrilinear $Z_{\mu}Z_{\nu}W^-_{\alpha}W^+_{\beta}$
gauge boson vertices that appear in these graphs are given,
respectively, by
\bea
\Gamma_{\mu\alpha\beta}(q,p_1,p_2) &=&   
(q-p_1)_{\beta}g_{\mu\alpha} + (p_1-p_2)_{\mu}g_{\alpha\beta} 
+(p_2-q)_{\alpha}g_{\beta\mu} \; ,
\label{G} \\
G_{\alpha\beta\mu\nu}(p_1,p_2,k_1,k_2)&=&  
2g_{\alpha\beta}g_{\mu\nu} - g_{\alpha\mu}g_{\beta\nu} 
                            - g_{\alpha\nu}g_{\beta\mu} \; ,
\label{4G}
\eea
where all momenta are considered incoming.

The case where one $Z$ is emitted from a  fermion line while the other
is emitted from a $W$ is described by the  eight graphs  denoted by
$T_{Wf}$:
\be
T_{Wf}= T_{1d}+T_{1u} + T_{2d}+T_{2u}
+T_{1\bar{d}}+T_{1\bar{u}} + T_{2\bar{d}}+T_{2\bar{u}}\; ,
\label{TWf}
\ee
and depicted in  Fig.~2.
These graphs contain one fermion propagator 
and one trilinear gauge boson  vertex. Only one of the currents
$V_{\alpha}$ or $U_{\beta}$  appears.   
A typical $T_{Wf}$ graph has the following structure:
\be
T_{1d}=
\frac{V_{\alpha}}{ W_1}\frac{1}{D_1}
\Gamma_{\alpha\rho\mu}(p_1,k_1-p_1,-k_1)
\ell_d\frac{g}{\sqrt{2}}\overline{u}_d\gamma_{\nu}P_L
\frac{1}{\qsl_2-(\ksl_1-\slp_1)}\gamma^{\rho}u_u
\epsilon_1^{\mu}\epsilon_2^{\nu} \; .
\label{1d}
\ee
 The couplings of the fermions to the $Z$ are defined here as 
\be
\frac{ig}{c_w}\gamma^{\mu}\left(\ell_f P_L +r_f P_R \right)  
\ee
where 
\be
\ell_f=I^3_{W,f}-s_w^2Q_f, ~~~~~~~~
r_f=-s_w^2Q_f ~. 
\ee
In the above expressions 
 $I^3_{W,f}$ is the third component of the weak isospin,  $Q_f$ the 
charge of fermion $f$, and   $P_{R,L}=(1 \pm \gamma_5)/2$. 

 The remainder of the $T$ graphs, where {\it both}
 $Z$s emanate from  fermions,  are denoted $T_{ff}$ and are exhibited in
Fig.~3:
 \be
T_{ff} = T_{W_1} + T_{W_2} + T_{D_1} + T_{D_2}\; .
\label{ff}
\ee
Each one of these graphs contains two fermion propagators.
In Eq.~(\ref{ff}) we have grouped
them according to the $W$ propagator they contain. 
In this class of graphs either one current appears and
two fermionic propagators are present in the other fermion line 
($T_{W_1}$ and $T_{W_2}$ graphs, 6 graphs each), or
no current appears and each fermion line contains one fermion propagator 
($T_{D_1}$ and $T_{D_2}$ graphs, 4 graphs each). 
Altogether there are 20 graphs in this class.

Finally, the $S$ graphs are divided into classes according to the
neutral gauge boson ($\gamma$, $Z$, or $g$) into which the initial quarks
anihilate. There are 63 graphs in this category, denoted as
$S_Z$ (20), $S_{\gamma}$ (20), $S_g$ (20), and $S_{ZZ}$ (3) according to the
gauge boson propagators that they contain. In the $S_V$ graphs
the $Z$s are emitted from fermions and thus all these graphs contain
 two fermion propagators. The $S_{ZZ}$ graphs are the Higgs-strahlung  graphs.
 Like the $W$ fusion graphs they contain no fermion propagators.  
One representative of each of these classes is shown in Fig.~4.

These seemingly different classes of graphs contain in fact a large
number of identical
terms. These common terms are responsible for the large unitarity 
cancellations that take place in the sum of all graphs. 
On the other hand, unitarity violation occurs
within individual subsets of graphs, e.g. $T_{WW}$,
as a result of incomplete cancellations among such terms. 
The common terms among the different classes of graphs arise when
 momentum factors in the graphs' numerators trigger one
of the following tree-level Ward identities:
\bea
k^{\mu}\gamma_{\mu}\equiv \ksl&=& (\ksl-\qsl_i)+\qsl_i 
=S^{-1}(k-q_i)-S^{-1}(q_i)
\label{kfer}  ,\\
q^{\mu}\Gamma_{\mu\alpha\beta}(q,p_1,p_2)&=& 
[p_2^2 g_{\alpha\beta} -p_{2 \alpha}p_{2\beta}] - 
[p_1^2 g_{\alpha\beta} -p_{1 \alpha}p_{1\beta}] 
= U^{-1}_{\alpha\beta}(p_2)_W-U^{-1}_{\alpha\beta}(p_1)_W ,~~~~~
\label{qG1}\\ 
k_1^{\mu}G_{\alpha\beta;\mu\nu }(p_1,p_2;k_1,k_2 ) &=&
\Gamma_{\alpha\beta\nu}(p_1,p_2+k_1,k_2)
 - \Gamma_{\alpha\beta\nu}(p_1+k_1,p_2,k_2)                
 \label{q4G}   ,\\
k^{\mu}_1 k^{\nu}_2g_{\mu\nu} &=& \frac{1}{2}(q^2-k_1^2-k_2^2)=
\frac{1}{2}D_H+\frac{1}{2}(M_H^2-2M_Z^2) ,
\label{WIHiggs}
\eea
where Eq.~(\ref{WIHiggs}) is the Ward identity of the $HZZ$ vertex with the two
$Z$ bosons on shell, while 
$U^{-1}_{\alpha\beta}(p)_W= (p^2-M_W^2)g_{\alpha\beta} -p_{\alpha}p_{\beta}$
is the inverse propagator of the $W$ in the unitary gauge.
The inverse propagators generated in this way 
cancel one of the propagators of the graph, resulting in a structure that
mimics the structure of a different class.   
For example,  
 the two classes of graphs $WW$ and $Wf$ 
 become identical in form  when
one of the $W_i$ propagators is removed in the
$T_{WW}$ graphs while at the same time 
 the fermion propagator is removed in 
the $T_{Wf}$ graphs. The remaining terms in the Ward identities 
mostly cancel when the particles involved are on shell.
As we have already noted,  
the momentum factors in the numerators
that trigger the above Ward identities are furnished  by the  trilinear
gauge boson vertices and the longitudinal polarization vectors 
of the external $Z$ bosons. 
    In fact current conservation, good high energy behaviour, the
  equivalence theorem and the cancellation of gauge parameters
all have their origins in these
  and other similar Ward identities satisfied by the tree-level
  vertices of the Standard Model.  
  Indeed a set of tree-level graphs is characterised as gauge invariant
   when  none of its parts can
    resemble the structure of a different  set of graphs
   by use of a Ward identity. 


 With this criterion in mind 
 it should  now be clear that the $S$ graphs are not involved in any
 unitarity cancellations with the $T$ graphs. First of all 
 the couplings involved are different. For example the $S_g$ graphs
 can never mix with any other class of graphs since they are the only
 ones that involve the  strong coupling constant. Thus they are gauge invariant
  and well behaved by themselves. Second,
  there are not enough factors  in  the numerators of these
  two classes  that are
  capable of cancelling the necessary  propagators to make a common
term.  One $W$ propagator will always survive in the $T$
   graphs while no such propagator is present in the $S$ graphs. 
   Thus in the search for a well behaved $W-$fusion squared amplitude
   the $S$ graphs can be ignored, and we need only concentrate
   on the $T$ graphs.
   
    With these considerations it should also 
    be evident that {\it off-shell} $Z-$fusion or Higgs-strahlung 
cannot give rise to
unitarity violation \cite{StiKle86}.
The Higgs-strahlung graphs $S_{ZZ}$ cannot communicate
     in any way with the rest of the graphs.
      Even if their Higgs 
       propagator is cancelled they will always contain two $Z$ boson 
        propagators.

 Equivalently the good high energy behaviour 
 of the $T$ graphs alone can also be verified by considering 
a different $W$-fusion process where the initial fermions cannot annihilate,
 like $e^+\mu^-\rightarrow \nu \bar{\nu}_{\mu} Z Z $. 
This process still retains the 
$T$ graphs but does not involve any $S$ graphs.   
\bigskip

 The  Ward identities of
 Eqs.~(\ref{kfer}-\ref{WIHiggs})
guarantee both the gauge parameter
independence of the
 amplitude as well as  independence with respect to gauge transformations
  of the external gauge bosons. For physical amplitudes with 
  external gauge bosons,  gauge invariance is encoded 
   in the following relationships:
   \bea
 \gamma ;g:~~~~~  k^{\mu} T_{\mu}(\gamma;g) &=&0 \label{wig}\\
 W^{\pm} ~:~~~~  k^{\mu} T_{\mu}(W^\pm)&=&\pm M_W {\cal T}(\phi) 
 \label{wiw} \\
 Z~:~~~~~~~k^{\mu} T_{\mu}(Z) &=& -iM_Z {\cal T}(\chi) 
 \label{wiz}  
   \eea 
 In these equations $T_{\mu}(\gamma;g),~T_{\mu}(W^\pm),~T_{\mu}(Z)$ 
 are physical amplitudes with at least one external gauge boson 
 $\gamma,~g,~W^{\pm},~Z$ carrying momentum $k^{\mu}$. The amplitudes 
   ${\cal T}(\phi)$ and ${\cal T}(\chi)$ on the right-hand side
are the corresponding
   set of graphs where the gauge boson is replaced by its would-be
    Goldstone boson. The above Eqs.~(\ref{wig}-\ref{wiz}) can be 
considered as Ward identities for whole set of graphs and they are the  
direct consequence of the Ward identities of 
 Eqs.(\ref{kfer}-\ref{WIHiggs})
 A set of graphs that satisfies these Ward identities
 is said to be gauge invariant. It would be both independent of any 
gauge parameters and of the choice of polarization vector for the external 
gauge boson. 
For the photon, Eq.~(\ref{wig}) is usually referred to 
as current conservation in QED,  
while from Eqs.~(\ref{wiw}) and (\ref{wiz}) the Equivalence Theorem 
follows directly \cite{Equiv}.

    In our case, since the fermions are massless and there is no $\chi WW$
    vertex, the $T_{Wf}$ and $T_{ff}$ graphs will give no contribution
    to the right-hand side of Eq.~(\ref{wiz}). Thus
\be
k_1^{\mu} T_{\mu}(Z)=-iM_Z{\cal T}_{WW}(\chi)\; .
 \ee   

    Since unitarity in {\it off-shell} $W-$fusion is violated by terms
    which are proportional to the {\it off-shellness} of the $W$ bosons
    \cite{StiKle86},
    the common terms between the $T_{WW}$ and the 
    $T_{Wf}+T_{ff}$ graphs will always be proportional to $W_1$ and $W_2$. 
    In the calculation of the squared amplitude 
    such common terms will emerge at three different levels. 
   The first level is that of the amplitude,
stripped of the polarization vectors of the $Z$s 
\be
T^{\mu\nu} =(T_{WW}+T_{Wf}+T_{ff})^{\mu\nu} \; .
\ee
At this level, the presence of the trilinear vertices already generates
{\it off-shell} 
unitarity violating terms in the $T_{WW}$ and  $T_{Wf}$ that cancel
between  them. At this stage all gauge parameters cancel also.

 The second cancellation occurs when in the full amplitude,
\be
T = T^{\mu\nu}\epsilon^{\mu}_1 \epsilon^{\nu}_2            \; ,
\ee
the polarization vectors of the $Z$ bosons become longitudinal and thus 
their leading part becomes proportional to the momenta of the $Z$s,
$k_1^{\mu}$ and $k_2^{\nu}$. 
Now the action of $k_1^{\mu}$ and/or $k_2^{\nu}$ on the amplitude will 
 produce extra {\it off-shell} terms.
 Such terms will again cancel  among the $T_{WW}$ 
 and  $T_{Wf}+T_{ff}$ graphs. 

 Finally, in the squared amplitude
 additional {\it off-shell} terms will appear between 
 the squared $W-$fusion amplitude $|T_{WW}|^2$
 and the interference term $T_{WW}(T_{Wf}^*+T_{ff}^*)$. 
 Most of these terms again cancel. Those that do not, 
  emerge from the interferences of the Higgs graph $T_H$. 
  Because of their structure (they contain no
fermion propagators), these terms will be allocated
  to the $W-$fusion amplitude $|T_{WW}|^2$.
 
 Accounting for  these cancellations at every level 
  is simply an exercise in identifing structures. The common terms
 can actually be identified graphically, as in Fig.~5 or Fig.~6.
 In order to keep track of them  the properties and the Ward identities 
 of the gauge boson  self-interaction vertices are exploited.
 Thus the trilinear gauge vertex of Eq.~(\ref{G}) is
  decomposed in the asymmetric form,
first used by t' Hooft \cite{Hooft},
\be
\Gamma_{\mu\alpha\beta}(q,p_1,p_2) =
\Gamma^F_{\mu\alpha\beta}(q;p_1,p_2)+\Gamma^P_{\mu\alpha\beta}(q;p_1,p_2)
\; ,
\label{FP}
\ee
where 
\bea
\Gamma^F_{\mu\alpha\beta}(q;p_1,p_2)&=& 
(p_1-p_2)_{\mu}g_{\alpha\beta}
-2q_{\alpha}g_{\beta\mu} +2q_{\beta}g_{\mu\alpha} \label{F} \\
\Gamma^P_{\mu\alpha\beta}(q;p_1,p_2)&=& 
-p_{1 \alpha}g_{\beta\mu} + p_{2 \beta}g_{\mu\alpha} \; .
\label{P}
\eea 
The pinch part is proportional to the momenta of two of
the legs of the vertex carrying their  corresponding Lorentz index. 
This part therefore vanishes identically when the two legs are on shell.
The $\Gamma^F$ part of the vertex satisfies the Feynman gauge part 
of the Ward identity of Eq.~(\ref{qG1}),
\be
q^{\mu}\Gamma^F_{\mu\alpha\beta}(q;p_1,p_2) = (p_2^2-p_1^2)g_{\alpha\beta}
=(W_2-W_1)g_{\alpha\beta}               \; ,
\label{qF}
\ee
while the pinch part $\Gamma^P$ will give rise to the longitudinal terms of 
Eq.~(\ref{qG1}).
The four-gauge-boson vertex of Eq.~(\ref{4G})
can also be split into two parts:
\bea
G_{\alpha\beta;\mu\nu}(p_1,p_2;-k_1,-k_2) &= &
G^F_{\alpha\beta;\mu\nu}(p_1,p_2;-k_1,-k_2)
+G^P_{\alpha\beta;\mu\nu}(p_1,p_2;-k_1,-k_2) \; ,\\ 
G^F_{\alpha\beta;\mu\nu}(p_1,p_2;-k_1,-k_2) &= &2g^{\alpha\beta}g^{\mu\nu}
\label{4GF} \; ,\\
G^P_{\alpha\beta;\mu\nu}(p_1,p_2;-k_1,-k_2)&= &
- g^{\alpha\mu}g^{\beta\nu} - g^{\alpha\nu}g^{\beta\mu}\; .
\label{4GP}
\eea
With the above decomposition,  the  Ward identity of Eq.(\ref{q4G}) 
is also satisfied individually by  its Feynman ($G^F$) and pinch ($G^P$)
parts. 

\bigskip

 We next proceed to explicitly exhibit and implement the aforementioned
cancellations between the $W-$fusion and the rest of the $T$ graphs.
 Most of the remaining part of this section will be 
  quite technical, involving extended use of the relevant Ward identities.
Readers who are only interested in the results of the calculation
may wish to go directly to
Eq.~(\ref{result}) and the discussion that follows it.
 
\vspace*{0.5cm}
\noindent{\underline{\it{The 1st cancellation}}}
\vspace*{0.3cm}

In order to identify the common {\it off-shell} terms at this level 
one must first decompose the trilinear and quadrilinear gauge vertices
into their Feynman and pinch parts.
The two sets of graphs $T_{WW}$ and $T_{Wf}$
which contain such vertices will then
separate into Feynman and pinch parts respectively:
\bea
T_{WW}^{\mu\nu}&=&T_{WW}^{F\mu\nu}+T_{WW}^{P\mu\nu}
\label{WWFP} \; ,\nonumber \\
T_{Wf}^{\mu\nu}&=&T_{Wf}^{F\mu\nu}+T_{Wf}^{P\mu\nu} \; .
\label{WfFP}
\eea
The pinch parts are proportional to the {\it off-shellness} of the $W$ bosons.
We will explicitly show that these parts exactly cancel.

Since the decomposition of a trilinear vertex in Eq.~(\ref{FP}) is
asymmetric, at this point one has to make a choice for the first
momentum argument of the $\Gamma^F$ vertices. Different choices should make 
no difference to the final result when the procedure is carried through
consistently. Nevertheless at intermediate points a particular choice may  
facilitate the calculations. It  turns out that choosing as the special
momentum for the $\Gamma^F$ vertices the momentum of 
 the corresponding off-shell $W$, $p_{1\alpha}$ or $p_{2\beta}$, 
 minimizes the number of interference terms that must be considered, thus
 expediting the calculations.\footnote{Actually this choice is
identical to the one
 made in the one-loop pinch technique, since these would be the momenta 
 outside the loop in $WW\rightarrow ZZ$.}
Thus the Feynman and pinch parts of the vertices are:
\be
\Gamma^F_{\alpha\rho\mu}(p_1;k_1-p_1,-k_1) = 
2(k_{1\alpha}g_{\rho\mu} 
+p_{1\mu}g_{\alpha\rho} -p_{1\rho}g_{\mu\alpha}) 
 ~;~~~
\Gamma^P_{\alpha\rho\mu}(p_1;k_1-p_1,-k_1)  =
(p_1-k_1)_{\rho}g_{\mu\alpha}\label{G11}
\ee
\be
\Gamma^F_{\beta\rho\nu}(p_2;k_2-p_2,-k_2)= 
2(k_{2\beta}g_{\rho\nu}
+p_{2\nu}g_{\beta\rho}-p_{2\rho}g_{\nu\beta} )~;~~~
\Gamma^P_{\beta\rho\nu}(p_2;k_2-p_2,-k_2) =
(p_2-k_2)_{\rho}g_{\nu\beta}
\label{G22}
\ee
\be
\Gamma^F_{\alpha\rho\nu}(p_1;k_2-p_1,-k_2) = 
2(k_{2\alpha}g_{\rho\nu} 
+p_{1\nu}g_{\alpha\rho} -p_{1\rho}g_{\nu\alpha}) 
~;~~~ 
\Gamma^P_{\alpha\rho\nu}(p_1;k_2-p_1,-k_2)  =
(p_1-k_2)_{\rho}g_{\nu\alpha}
\label{G12}
\ee 
\be
\Gamma^F_{\beta\rho\mu}(p_2;k_1-p_2,-k_1)= 
2(k_{1\beta}g_{\rho\mu}
+p_{2\mu}g_{\beta\rho}-p_{2\rho}g_{\mu\beta} )~;~~~
\Gamma^P_{\beta\rho\mu}(p_2;k_1-p_2,-k_1) =
(p_2-k_1)_{\rho}g_{\mu\beta}
\label{G21}
\ee  
 In the above expressions we have dropped all terms
 that will not contribute to the matrix element. These 
 are the terms proportional 
 to $p_{1\alpha}$ and $p_{2\beta}$, which for massless
 fermions vanish when contracted with the external
fermionic currents, i.e. $p_{1\alpha}V^{\alpha}=0$ and $p_{2\beta}U^{\beta}=0$.
 Furthermore at this level we can  
also drop terms proportional to $k_{1\mu}$ and $k_{2\nu}$ that
vanish when contracted with  the polarization vectors of the $Z$s:
$k_{1\mu}\epsilon^{\mu}(k_1)=0=k_{2\nu}\epsilon^{\nu}(k_2)$. 

It is straightforward to calculate the pinch part of the $T_{Wf}$ graphs.  
Since they contain only one trilinear vertex they split immediately into
two parts.  
One readily observes that the pinch part, $\Gamma^P$, of the
vertex simply cancels the fermion propagator in each
of these graphs by virtue of the Ward identity of Eq.~(\ref{kfer}).
 We illustrate this explicitly for one of the graphs.
The expression for  $T_{1d}$ of Fig.~2 has already been written out
 explicitly in  Eq.~(\ref{1d}).
Using Eq.~(\ref{G11}) for $\Gamma^P_{\beta\rho\nu}$,
the pinch part of this  graph  is  given by
\be
T_{1d}^P=\frac{V^{\alpha}}{ W_1}
\frac{g_{\alpha\mu}}{D_1}\frac{g}{\sqrt{2}}\ell_d
\overline{u}_d\gamma_{\nu}
\frac{1}{\qsl_2-\ksl_1+\slp_1}
(\slp_1-\ksl_1)P_L u_u(q_2)\epsilon_1^{\mu}\epsilon_2^{\nu} \; .
\label{qfer1d}
\ee
Then writing 
\be
\slp_1-\ksl_1=(\qsl_2-\ksl_1+\slp_1) - \qsl_2    \; ,
\ee
the first term will cancel the fermion propagator 
while the second one vanishes since $\qsl_2u_u(q_2)=0$.
So the pinch part of this graph is 
\be
T_{1d}^P
=\frac{V^{\alpha}}{ W_1}\frac{U^{\beta}}{W_2}(\ell_d)
\frac{W_2}{D_1}g_{\alpha\mu}g_{\beta\nu}\epsilon_1^{\mu}\epsilon_2^{\nu}\;
.
\label{T1dP}
\ee

The corresponding expression $T_{1u}^P$, of  graph $T_{1u}$ of Fig.~2
is simply obtained by (i)  changing $\ell_d$, 
the fermion coupling of the $Z$, to $\ell_u$  
since the $Z$ boson is now emitted from the
$up$ fermion, and (ii)  changing the overall sign,
since now the fermion propagator involves $\qsl_3-(\slp_1-\ksl_1)$.
Thus, using the following relation for the fermion couplings,  
\be
\ell_u-\ell_d=c_w^2              \; ,
\ee
the two graphs combined give 
\be
T^P_1\equiv T_{1u}^P+T_{1d}^P=
\frac{V_{\alpha}}{ W_1}\frac{U_{\beta}}{W_2}(-c_w^2)
\frac{W_2}{D_1}g_{\alpha\mu}g_{\beta\nu} \; .
\label{1P}
\ee
Evidently this expression
resembles the structure of the $W-$fusion graphs,
$T_{WW}$. It contains no fermion propagators and has 
the correct  coupling. We have divided and multiplied by the
$W$ boson inverse propagator $W_2$ in order to bring the expression into
the form of Eq.~(\ref{WWstruct}). This step is not strictly necessary,
 but our convention will be always to  
 extract a factor $V_{\alpha}U_{\beta}/(W_1W_2) $ from all such terms.

The pinch parts of the rest of the $T_{Wf}$ graphs
are extracted in a similar way.
The graphs always combine in pairs to produce the correct  coupling $c_w^2$.
Altogether we obtain for the pinch terms of the $T_{Wf}$ graphs 
the following expression:  
\be
T_{Wf}^P=\frac{V_{\alpha}}{ W_1}\frac{U_{\beta}}{W_2}(-c_w^2)
(W_1+W_2)\{\frac{g_{\alpha\mu}g_{\beta\nu}}{D_1}
+\frac{g_{\alpha\nu}g_{\beta\mu}}{D_2}
\}\epsilon_1^{\mu}\epsilon_2^{\nu} \; .
\label{WfP}
\ee

\bigskip

 Next we turn to the $W-$fusion graphs.
In order to identify similar pinch terms  
to those of the $T_{Wf}$ graphs we will use the
following identity for the product of two trilinear vertices
\cite{CoPa89}:
\be
\Gamma_{\alpha\rho\mu}\Gamma_{\beta\rho\nu}= 
\Gamma^F_{\alpha\rho\mu}\Gamma^F_{\beta\rho\nu}
+\Gamma^P_{\alpha\rho\mu}\Gamma_{\beta\rho\nu}
+\Gamma_{\alpha\rho\mu}\Gamma^P_{\beta\rho\nu}
- \Gamma^P_{\alpha\rho\mu}\Gamma^P_{\beta\rho\nu} \; .
\label{GG}
\ee
This decomposition will enable us to make use of the 
Ward identity of the full vertex given in  Eq.~(\ref{qG1}).

Let us first consider the graph $T_W$ given in Eq.~(\ref{W}).
Using Eq.~(\ref{GG}), the product of its trilinear vertices
can be expressed as 
\bea
\Gamma_{\alpha\rho\mu}(p_1,k_1-p_1,-k_1)
\Gamma_{\beta\rho\nu}(p_2,k_2-p_2,-k_2) &=&
\Gamma^F_{\alpha\rho\mu}(p_1;k_1-p_1,-k_1)
\Gamma^F_{\beta\rho\nu}(p_2;k_2-p_2,-k_2) \nonumber \\
&+&
g_{\beta\nu}(p_2-k_2)_{\rho}\Gamma^{\alpha\rho\mu}(p_1,k_1-p_1,-k_1)
\nonumber \\
 &+&g_{\alpha\mu}(p_1-k_1)_{\rho}\Gamma^{\beta\rho\nu}(p_2,k_2-p_2,-k_2)
\nonumber \\ 
&-&g_{\alpha\mu}g_{\beta\nu}(p_2-k_2)\cdot (p_1-k_1) \; .
\eea
Since  $p_1-k_1=k_2-p_2$,  the Ward identity of Eq.~(\ref{qG1})
is immediately triggered 
by the second and third term of the above equation. This will produce
{\it off-shell} terms proportional to
$p_i^2=W_i+M_W^2$. The fourth term
can be written in terms of the inverse propagator $D_1$ when $M_W^2$ 
is added and subsequently subtracted.    
Finally, using 
the on-shell condition for the $Z$s, $k_1^2=k_2^2=M_Z^2$,
 the expression (\ref{W}) for the diagram $T_W$
  is transformed into 
\be
T_{W}^{\alpha\beta\mu\nu}=\frac{c_w^2}{D_1}\left[
\Gamma^F_{\alpha\rho\mu}\Gamma^F_{\beta\rho\nu} 
+[ M_Z^2(c_w^2-2s_w^2) + D_1 + (W_1+W_2) ]g_{\alpha\mu}g_{\beta\nu}
\right]         \; .
\label{WP}
\ee
The momenta arguments of the vertices will no longer be exhibited, since 
the way their momenta are assigned  should be 
 obvious from the Lorentz indices of the vertices.
The first and third indices determine  the first and third 
momentum arguments of the vertex respectively, according to 
$(\mu,\nu,\alpha,\beta) \rightarrow (-k_1,-k_2,p_1,p_2)$, while the 
second argument is fixed by momentum conservation. 
In an identical way the crossed graph $T_{cW}$ is rewritten as 
\be
T_{cW}^{\alpha\beta\mu\nu}=\frac{c_w^2}{D_2}\left[
\Gamma^F_{\alpha\rho\nu}\Gamma^F_{\beta\rho\mu} 
+[ M_Z^2(c_w^2-2s_w^2) + D_2 + (W_1+W_2) ]g_{\alpha\nu}g_{\beta\mu}
\right]
\label{cWP}
\ee

Having obtained  these new forms for the 
 diagrams $T_{W}$ and $T_{cW}$ we make 
the following observations. The terms proportional to 
$M_Z^2(c_w^2-2s_w^2)$,  
directly  combine with 
the corresponding  would-be Goldstone graphs
of Eq.~(\ref{f}) and Eq.~(\ref{cf}) to  produce an overall coupling
equal to  
$ s_w^4-2s_w^2c_w^2+c_w^4=(c_w^2-s_w^2)^2 $. 
The  terms proportional to
$D_1$ or $D_2$ will immediately cancel the relevant propagator of 
the graph, i.e. either $1/D_1$ or $1/D_2$, and will thus combine with the 
quadrilinear vertex graph $T_{4G}$. 
  In doing so they cancel the $G^P$ pinch part of the
quadrilinear vertex.  
The last terms proportional to the {\it off-shellness} of the $W$ bosons 
are the pinch terms.
These operations are 
represented pictorially for graph $T_{W}$ in Fig.~\ref{fig5}(a),   
where the ellipsis represent the $M_Z^2(c_w^2-2s_w^2)$ term of 
Eq.~(\ref{WP})

Thus the Feynman and pinch parts of the  $T_{WW}$ graphs are 
given by the following expressions: 
\bea 
T^{F\mu\nu}_{WW}&= &\frac{V_{\alpha}}{ W_1}\frac{U_{\beta}}{W_2}
\Big\{ 
\frac{c^2_w}{D_1}\Gamma^F_{\alpha\rho\mu}\Gamma^F_{\beta\rho\nu} 
+\frac{c^2_w}{D_2}\Gamma^F_{\alpha\rho\nu}\Gamma^F_{\beta\rho\mu}
+2c_w^2g_{\alpha\beta}g_{\mu\nu} ~~~~~~~~~~
\nonumber \\
&&
~~~~~~~~~~+M_Z^2(c_w^2-s_w^2)^2[\frac{g_{\alpha\mu}g_{\beta\nu}}{D_1}
+\frac{g_{\alpha\nu}g_{\beta\mu}}{D_2}]  
+M_Z^2\frac{g_{\alpha\beta}g_{\mu\nu}}{D_H}
\Big\}\; ,
\label{WWF}
\eea
and 
\be
T^{P\mu\nu}_{WW}= \frac{V_{\alpha}}{W_1}\frac{U_{\beta}}{W_2}
(c^2_w)(W_1+W_2)
\{\frac{g_{\alpha\mu}g_{\beta\nu}}{D_1}
+\frac{g_{\alpha\nu}g_{\beta\mu}}{D_2}
\}\; .
\label{WWP}
\ee
We note that the {\it off-shell} pinch terms have only been
generated from the graphs that contain trilinear gauge vertices, namely
$T_{W}$ and $T_{cW}$. Only these two graphs can communicate with the 
rest of the graphs, in this case $T_{Wf}$, due to the rich momentum 
structure of their numerator.

Finally we observe that the pinch part  
of the $W-$fusion graphs, $T^{P\mu\nu}_{WW}$ in the above equation,
exactly cancels the pinch term 
of the $T_{Wf}$ graphs $T^P_{Wf}$ of Eq.~(\ref{WfP}).
Indeed 
\be
T^P_{WW}+T^P_{Wf}=0 \; .
\ee

We have also verified the above results by performing the calculations 
 in a general $R_{\xi}$ gauge. 
 For this $2 \to 4$ process the cancellation of the gauge parameters 
 due to current conservation is not automatic as in the case 
 of a $2 \to 2$ process. Common pinch terms among
 different classes of graphs are now even more prolific, 
    due to the presence of the extra momenta in the longitudinal part 
   of the gauge bosons' propagators. 
   In addition, most of them will also be gauge parameter dependent.
   When all such terms are identified and 
   cancelled the surviving expressions coincide with 
   those obtained in the Feynman gauge.
    Thus the formula of Eq.~(\ref{WWF}) for $T_{WW}^F$ and the corresponding
    ones for $T_{Wf}^F$ and $T_{ff}$ are truly gauge parameter independent 
    expressions. 
    
    It would be interesting to find or invent a gauge where these expressions
     could be obtained {\it automatically} by the Feynman rules of the 
     particular gauge. In such a gauge this first level 
      of cancellations would be avoided. Furthermore such a gauge 
      might prove useful in other multi-gauge-boson processes. 
      However we are not aware of any such special gauge.
        Although the $\Gamma^F$ vertices of Eqs.~(\ref{G11}--\ref{G21}) 
    look similar to the trilinear vertices of the Gervais-Neveu 
    gauge \cite{CN},  
    they are actually not the same, and thus the
     Gervais-Neveu  gauge expression of 
    the $W-$fusion graphs does not coincide with $T_{WW}^F$.
        
     Although the expression for the $W$ fusion graphs Eq.(\ref{WWF}), 
     obtained after this first step, is gauge 
     parameter independent,  it remains however gauge non-invariant 
     in the sense that it still does not satisfy the Ward identity of 
     Eq.~(\ref{wiz}). Because of this fact, the $W$ fusion part can still not
     be separated from the rest of the graphs. To do this, the cancellations 
     inherent in Eq.~(\ref{wiz}) must be allowed to 
     take place first. This is done in the next step. 
We also note that the following steps would not be necessary had we 
considered photons instead of $Z$s in the final state, i.e.  
$WW\rightarrow \gamma\gamma$. 

\vspace*{0.5cm}
\noindent{\underline{\it{The 2nd cancellation}}}
\vspace*{0.3cm}

After this first cancellation has taken place  
the resulting gauge parameter independent amplitude $T$ for our 
process can be written as 
\bea
T &=& (T^F_{WW}+T^P_{WW} + T^F_{Wf}+T^P_{Wf} + T_{ff}) ^{\mu\nu}
\epsilon^{\mu}_1 \epsilon^{\nu}_2 \nonumber \\
&=& (T^F_{WW} + T^F_{Wf} + T_{ff})^{\mu\nu}
\epsilon^{\mu}_1 \epsilon^{\nu}_2 ~. 
\eea

The further cancellations due to the longitudinal $Z$ bosons 
become more transparent in the
squared unpolarized amplitude where  
the sum over 
the polarizations of the $Z$s will  give a factor
\be
\sum_{\lambda_1}\epsilon^{\mu}_Z(k_1,\lambda_1) 
   \epsilon^{\ast\mu '}_Z(k_1,\lambda_1) =
   -g_{\mu\mu '}+\frac{k_1^{\mu}k_1^{\mu '}}{M_Z^2} \; ,
\ee
and similarly for  $\epsilon^{\nu}_Z(k_2,\lambda_2)$.  
The extra momentum factors, $k_1^{\mu},~k_2^{\nu}$ etc. 
introduced in this way  will result in bad 
high energy behaviour,  since their growth with energy 
  cannot be compensated by the constant factors of  $M_Z$ 
  in the denominator.  
 Thus cancellations at this stage will be instrumental 
 in  restoring unitarity.

The unpolarized squared amplitude is given by
\be
\overline{~|T|^2}\!= \!\frac{1}{4}
T^{F\mu\nu}\!
\left(-g_{\mu\mu '}+k_{1\mu}k_{1\mu '}/M_Z^2\right)\!
\left(-g_{\nu\nu '}+k_{2\nu}k_{2\nu '}/M_Z^2\right)\!
T^{F\ast}_{\mu '\nu '} \; .
\label{fullamp}
\ee
Expanding the product of polarization tensors gives four terms:  
\bea
4\overline{~|T|^2}&= & T^{F\mu\nu}T^{F\ast}_{\mu\nu}  
                  -\frac{k_{1\mu}}{M_Z}T^{F\mu\nu}
                     \frac{k_{1\mu'}}{M_Z}T^{F\ast}_{\mu'\nu}
\nonumber \\
 && -\frac{k_{2\nu}}{M_Z}T^{F\mu\nu}
                     \frac{k_{2\nu'}}{M_Z}T^{F\ast}_{\mu\nu'}  
+\frac{k_{2\nu}k_{1\mu}}{M^2_Z}T^{F\mu\nu}
\frac{k_{2\nu'}k_{1\mu'}}{M^2_Z}T^{F\ast}_{\mu'\nu'} \; .
\label{fullamp4}
\eea
We next determine the effect of the factors of longitudinal 
momenta within each of the following terms: 
\be
k_{1\mu} T^{F\mu\nu},~~~~
k_{2\nu} T^{F\mu\nu},~~~~
\mbox{and}~~
k_{2\nu}k_{1\mu}T^{F\mu\nu}\; ,
\ee 
before actually squaring them.  We will then explicitly show 
 the generation and cancellation of the 
common  {\it off-shell} terms 
within each of the above terms.  

The action of  
 $k_1^{\mu}$ on the $T_{WW}$ amplitudes will generate 
{\it off-shell} terms when contracted with the $\Gamma^F$ vertices. 
Since $k_1^{\mu}$ is not the first, special, argument of $\Gamma^F$ the 
Feynman Ward identity of Eq.~(\ref{qF}) is modified to:
\be
k_1^{\mu} \Gamma^F_{\alpha\rho\mu}= 
\left( -D_1 + M_Z^2 + W_1 \right)g_{\alpha\rho} 
+2k_{1\alpha}(k_1-p_1)_{\rho} \label{k1G1} 
\ee
for the graph $T_W$, and to 
\be
k_1^{\mu}\Gamma^F_{\beta\rho\mu}= 
\left( -D_2 + M_Z^2 + W_2 \right)g_{\beta\rho} 
+ 2k_{1\beta}(k_1-p_2)_{\rho} \label{k1G2}
\ee
for the crossed graph $T_{cW}$. 
The remaining terms $(k_1-p_1)_{\rho}$ and $(k_1-p_2)_{\rho}$ 
of the above equations will create additional 
{\it off-shell} terms when they act in turn on the remaining $\Gamma^F$ 
vertex of each graph. Now the modified Ward identities read:
\bea
(k_1-p_1)^{\rho}\Gamma^F_{\beta\rho\nu} =&
-[D_1+W_2+M_Z^2(c_w^2-s_w^2)]g_{\beta\nu}-2k_{2\beta}k_{2\nu} \; ,\nonumber \\
(k_1-p_2)^{\rho}\Gamma^F_{\alpha\rho\nu}=&
-[D_2+W_1+M_Z^2(c_w^2-s_w^2)]g_{\alpha\nu}-2k_{2\alpha}k_{2\nu} \; ,
\eea
where the following elementary identities have been used for the dot products:
\bea 
 2 k_1\cdot p_1&=&  -D_1 + M_Z^2 + W_1\; ,~~~ 
 2 k_1\cdot p_2=  -D_2 + M_Z^2 + W_2  \; ,\\
2 k_2\cdot p_1 &= & -D_2 + M_Z^2 + W_1\; ,~~~  
2 k_2\cdot p_2 =   -D_1 + M_Z^2 + W_2 \; .
\eea
In the end  the contraction of $k_1^{\mu}$ with the 
$T^F_{WW}$ graphs assumes the following form: 
\be
k_1^{\mu}T^{F\mu\nu}_{WW} = 
M_Z^2A_1^{\nu} 
-2c_w^2 F_1 k_2^{\nu}+M_Z^2 \frac{F_2}{D_H} k_1^{\nu}
+ c_w^2 O_1^{\nu} ~.  
\label{k1TWW}
\ee
In the term $A_1^{\nu}$ the index  $\nu$ is never 
carried by either $k_2$ or $k_1$. The dependence on 
$k_2^{\nu}$ or $k_1^{\nu}$ has been explicitly exhibited,
the latter emerging from the Higgs graph. The explicit expressions read:
\bea
A_1^{\nu}&=&
\frac{V^{\alpha}}{W_1}\frac{U^{\beta}}{W_2}
\left[\frac{c_w^2}{D_1}\Gamma^F_{\beta\alpha\nu}
 +\frac{c_w^2}{D_2}\Gamma^F_{\alpha\beta\nu} - (2c_w^2-1)
\left(  \frac{k_1^{\alpha}g^{\beta\nu}}{D_1} 
       +\frac{k_1^{\beta}g^{\alpha\nu}}{D_2} \right)\right]
\label{A1} \\
F_1 &=& \frac{V^{\alpha}}{W_1}\frac{U^{\beta}}{W_2}t^{\alpha\beta}
\label{F1} \\
F_2&=&\frac{V^{\alpha}}{W_1}\frac{U^{\beta}}{W_2}g^{\alpha\beta}
\label{H}
\eea
with 
\be
t^{\alpha\beta}=g^{\alpha\beta}+\frac{2k_1^{\alpha}k_2^{\beta}}{D_1} + 
\frac{2k_1^{\beta}k_2^{\alpha}}{D_2} \; .
\label{t}
\ee
The {\it off-shell}  terms $O_1$ in (\ref{k1TWW}) are 
\be
O_1^{\nu} = \frac{V^{\alpha}}{W_1}\frac{U^{\beta}}{W_2}
\left[
\frac{W_1}{D_1}\Gamma^F_{\beta\alpha\nu}
 +\frac{W_2}{D_2}\Gamma^F_{\alpha\beta\nu} 
 - 2\frac{W_2}{D_1} k_1^{\alpha}g^{\beta\nu}
 - 2 \frac{W_1}{D_2}k_1^{\beta}g^{\alpha\nu}\right] \; .
\ee
The action of $k_1^{\mu}$ on the $T_{Wf}$
 and $T_{ff}$ graphs 
 can be easily determined. Since mostly the $Z$ bosons are 
 emitted from fermion lines, the  $k_1^{\mu}$ will directly pinch 
 the fermion propagators
 adjacent to the $Z$ boson leg by virtue of the Ward identity 
 of Eq.~(\ref{kfer})
  with $k \rightarrow  k_1$. 

 When $k_1^{\mu}$  is emitted from a $W$ then the modified
 Ward identities of Eqs.~(\ref{k1G1}) and (\ref{k1G2}) 
 must be used. 
 The terms $(k_1-p_1)_{\rho}$ and $(k_1-p_2)_{\rho}$ 
 emerging from these identities will now pinch the fermion 
 propagator attached to the internal $W$ boson the same way as in 
 Eq.~(\ref{qfer1d}).
 From the remaining terms of  equations (\ref{k1G1}) and (\ref{k1G2})
 the ones proportional to $M_Z^2$ cannot obviously pinch while the rest 
 pinch one of the $W$ propagators of the graphs. 
 These latter pinch terms attain therefore a structure similar to that of the 
 $T_{ff}$ graphs and indeed will cancel  exactly the entire contribution 
 of the  $T_{ff}$ graphs. 

After carrying out the above steps  we finally obtain 
\be
k_1^{\mu}T^{F\mu\nu}_{Wf} = -c_w^2 O_1^{\nu} + M_Z^2 R_1^{\nu}
 - k_1^{\mu}T^{\mu\nu}_{ff} \; .
\label{k1TWf}
\ee
One immediately observes that 
the {\it off-shell} pinch terms that emerge from the $T_{Wf}$ graphs 
 are exactly opposite to the ones generated by the $W-$fusion graphs
 of Eq.~(\ref{k1TWW}) and 
  exactly cancel.
   This is a manifestation of the Ward identity of 
  Eq.~(\ref{wiz}).
 The remainder $R_1^{\nu}$  retains the structure 
 of the $T_{Wf}$ graphs and is given explictly by 
 \bea
R_1^{\nu}&=& \frac{V_{\alpha}}{ W_1}\frac{1}{D_1}
\frac{g}{\sqrt{2}}
\overline{u}(q_4) \left\{\ell_d\gamma^{\nu}
\frac{1}{\qsl_4+\ksl_2}\gamma^{\alpha}P_L
+
\ell_u\gamma^{\alpha}P_L
\frac{1}{\qsl_2-\ksl_2}\gamma^{\nu}
\right\}u(q_2)\nonumber \\
&+& \frac{U_{\beta}}{ W_2}\frac{1}{D_2}
\frac{g}{\sqrt{2}}
\overline{v}(q_1) \left\{\ell_d\gamma^{\beta}P_L
\frac{1}{\qsl_3+\ksl_2}\gamma^{\nu}
+\ell_u\gamma^{\nu}\frac{1}{\qsl_1-\ksl_2}\gamma^{\beta}
P_L\right\}v(q_2)  \; .
\label{R1}
 \eea
Thus the action of $k_1^{\mu}$ on the complete amplitude 
will finally be of the form:
 \be
k_{1\mu}T^{\mu\nu}= M_Z^2A_1^{\nu} 
-2c_w^2 F_1 k_2^{\nu}+M_Z^2\frac{F_2}{D_H} k_1^{\nu} 
+M_Z^2 R_1^{\nu} \; .
\label{k1T}
\ee
As already mentioned,
this is nothing else but the Ward identity of Eq.~(\ref{wiz}) slightly modified
by the remainder of the $T_{Wf}$ graphs. The fact that such a term
 has survived 
in the left-hand side is simply because after the first cancellation 
the amplitudes have been written in terms of modified vertices $\Gamma^F$.
Since we also do not contract with $\epsilon^{\nu}(k_2)$,   the terms 
proportional to $k_2^{\nu}$ have also survived. 

\bigskip

The effect of the longitudinal factor $k_2^{\nu}$ 
in the third term of Eq.~(\ref{fullamp4}) is obtained in
an identical manner. 
As expected, the {\it off-shell} terms
common between the $T_{WW}$ and $T_{Wf}$
graphs again  cancel. In the end one obtains
 \be
k_{2\nu}T^{\mu\nu}= M_Z^2A_2^{\mu} 
-2c_w^2 F_1 k_1^{\mu}+M_Z^2 \frac{F_2}{D_H} k_2^{\mu} 
+ M_Z^2 R_2^{\mu} \; ,
\ee
where $A_2^{\mu}$ and $R_2^{\mu}$ can be obtained from
Eqs.~(\ref{A1}) and (\ref{R1}) respectively by
the replacements
$k_1 \leftrightarrow k_2$ and $\mu \leftrightarrow \nu$.

Finally we examine the fourth term of Eq.~(\ref{fullamp4}),
$k_{2\nu}k_{1\mu} T^{\mu\nu}$. Here
the momenta of both $Z$s act on the amplitude.
(The order in which the  contractions are performed does not matter.)
Acting   
first with $k_1^{\mu}$  again  produces the result
of Eq.~(\ref{k1T}).
When next $k_2^{\nu}$ is contracted with this equation it
generates {\it off-shell} terms from both classes of graphs.
On the $T_{WW}$ graphs it gives
\be
k_2^{\nu}k_1^{\mu}T^{F\alpha\beta\mu\nu}_{WW} = M_Z^2\left[
 (\frac{1}{2}-4c_w^2) F_1 +   A F_2+
 c_w^2O_{12}F_2\right]
\label{k2k1TWW}
\ee
with 
\be
A=
\left[\frac{M_W^2}{D_1}+\frac{M_W^2}{D_2}+\frac{M_H^2-2M_Z^2}{2D_H}\right]
\ee
and
\be
O_{12}= 
\left[\frac{W_2}{D_1}+\frac{W_1}{D_2}\right] \; .
\ee
Here we have also used the Ward identity of the 
Higgs vertex, Eq.~(\ref{WIHiggs}),
to extract a piece from
the Higgs graph and combine it with the gauge boson graphs in $F_1$.

The result for the remainder of the  $T_{Wf}$ graphs $R_1^{\nu}$ 
can be readily obtained
by removing the fermionic propagators  in Eq.~(\ref{R1}) with $\ksl_2$,
by virtue of the Ward identity of Eq.~(\ref{kfer}).
 This gives   
\bea
k_2^{\nu}R_1^{\nu} &=&- c_w^2 O_{12}\; .
\label{k2k1TWf}
\eea
Finally collecting together Eqs.~(\ref{k2k1TWW}) and  (\ref{k2k1TWf})
we obtain
\be
k_2^{\nu}k_1^{\mu}T^{\mu\nu}=  M_Z^2\left[
(\frac{1}{2}-4c_w^2) F_1 + F_2 A \right] \; .
\label{k2k1T}
\ee
Notice  again that all {\it off-shell} terms have cancelled.

\vspace*{0.5cm}
\noindent{\underline{\it{The 3rd cancellation}}} 
\vspace*{0.3cm}

 After the first two cancellations have taken place,  the 
 unpolarized squared amplitude of Eq.~(\ref{fullamp4}) attains the following
 form 
\bea
4\overline{~|T|^2}&= & T^{F\mu\nu}T^{F\ast}_{\mu\nu}
\nonumber \\
&&-\frac{1}{M_Z^2}\left[ M_Z^2A_1^{\nu} -2c_w^2 F_1 k_2^{\nu}
+F_2\frac{M_Z^2}{D_H} k_1^{\nu}
+M_Z^2 R_1^{\nu} \right] 
\nonumber \\ 
&&~~~~~~~\times \left[ M_Z^2A_1^{\nu} -2c_w^2 F_1 k_2^{\nu}
+ F_2\frac{M_Z^2}{D_H}k_1^{\nu}
+M_Z^2 R_1^{\nu} \right]^{\ast}
\nonumber \\
&&-\frac{1}{M_Z^2}\left[ M_Z^2A_2^{\mu} -2c_w^2F_1  k_1^{\mu}
+F_2\frac{M_Z^2}{D_H} k_2^{\mu}
+M_Z^2R_2^{\mu}  \right] 
\nonumber \\ 
&&~~~~~~~\times 
\left[ M_Z^2A_2^{\mu} -2c_w^2 F_1 k_1^{\mu}+
 F_2\frac{M_Z^2}{D_H} k_2^{\mu}
+M_Z^2R_2^{\mu}  \right]^{\ast}
\nonumber \\
&&+ \left\vert(\frac{1}{2}-4c_w^2) F_1 + F_2 A\right\vert^2 \; .
\label{3rd}
\eea

We can now see the advantages of our choice for the  $\Gamma^F$ vertices.
When each term of Eq.~(\ref{3rd}) is squared, additional {\it off-shell}
terms will appear due to the interference of the $T_{WW}$ with the 
$T_{Wf}+T_{ff}$ graphs. This comes about from the action of the 
terms $k_1^{\mu}$, $k_2^{\nu}$,  $k_1^{\nu}$ and $k_2^{\mu}$ which 
have been written explicitly in Eq.~(\ref{3rd}). No such terms
will be generated in the first term however. 
Squaring the second term and exhibiting only
terms that are of the $W-$fusion type we obtain
\bea
& & -\frac{1}{M_Z^2}\Bigg[ ~M_Z^4 A_1\cdot A_1^* + 4c_w^4 M_Z^2 |F_1|^2
+ M_Z^2  |F_2/D_H|^2 ~~~~~~~~~~~~\nonumber \\ 
& & ~~~~~~~~~~-2c_w^2 F_1 k_2^{\nu}
\left( M_Z^2 A_1^{\nu}+M_Z^2F_2\frac{M_Z^2}{D_H} k_1^{\nu}
+M_Z^2R_1^{\nu}  \right)^* + c.c. 
\nonumber \\ 
 & & ~~~~~~~~~~ + F_2\frac{M_Z^2}{D_H}
 k_1^{\nu}
\left( M_Z^2 A_1^{\nu}+M_Z^2 R_1^{\nu}  \right)^*
 + c.c. +... 
\Bigg]~~~~ \; .
\label{square}
\eea
On the second line of this expression
(\ref{square}),
$k_2^{\nu}$ will generate 
the same {\it off-shell} terms, $O_{12}$,  
 in a manner identical to that of the previous 
subsection. These terms will cancel in the same way they did in
Eqs.~(\ref{k2k1TWW}) and (\ref{k2k1TWf}).
Thus the second line of (\ref{square}) is equal to
\be
-2c_w^2 F_1 M_Z^2\Big[  (\frac{1}{2}-2c_w^2) F_1^* +  A^* F_2^*\Big]\; .
\ee
Next we consider the third line of Eq.~(\ref{square}).
This corresponds to the interference of the contracted Higgs graph
$k_1^{\mu}T_H^{\mu\nu}$ with whatever has remained. 
The index $\nu$ is now carried by the four momentum $k_1$ instead of $k_2$. 
This will not affect the {\it off-shell} terms generated by the $W$ 
fusion graphs which will be again $O_{12}$, 
\be
M_Z^2k_1^{\nu}A_1^{\nu}=M_Z^2h_1+M_Z^2c_w^2O_{12}F_2 \; ,
\label{k1A1}
\ee
with 
\bea
h_1&=&\frac{V^{\alpha}}{W_1}\frac{U^{\beta}}{W_2}
\Bigg[
-c_w^2g_{\alpha\beta}\left( \frac{D_2}{D_1}+ \frac{D_1}{D_2}\right)
+2c_w^2(k_{1\alpha}k_{2\beta}-k_{2\alpha}k_{1\beta})
\left( \frac{1}{D_1}- \frac{1}{D_2}\right)
\nonumber \\
&&
+\left([g_{\alpha\beta}M_W^2
+(1-4c_w^2)k_{1\alpha}k_{1\beta}\right]
\left( \frac{1}{D_1}+ \frac{1}{D_2}\right)
\Bigg] \; .
\label{h1}
\eea
On the other hand, $k_1^{\nu}$ acting on the 
 remainder of the $T_{Wf}$ graphs $R_1^{\nu}$ 
 now generates {\it off-shell} terms with the opposite sign from
 the ones that would be generated from $k_2^{\nu}$. This can easily be seen 
 by looking at the expression for $R_1^{\nu}$ in Eq.~(\ref{R1}).
 For $\ksl_1$ to pinch the propagator $1/(\qsl_4+\ksl_2)$ we must 
 use momentum conservation to write 
 $\ksl_1=-\ksl_2+\slp_1+\slp_2$, thus introducing a minus sign and a
 remainder proportional to $\slp_1+\slp_2=\ksl_1+\ksl_2$ which can give
 zero pinch. 
 The final result will assume the form   
 \be
k_1^{\nu}R_1^{\nu}  = 
 c_w^2M_Z^2O_{12}F_2 \; .
\label{k1k1TWf}
 \ee
 Thus we observe that the  {\it off-shell} terms in this case do not cancel 
 between Eqs.~(\ref{k1A1}) and (\ref{k1k1TWf}).
 The third line of (\ref{square}) is then equal to
\be 
F_2\frac{M_Z^2}{D_H}\left[ M_Z^2h_1^* + 2\cdot c_w^2M_Z^2O_{12}F_2^*  
+ ....\right]       \; .
\ee 
 Since the {\it off-shell} terms that have survived in this
  interference of the Higgs graph do not contain any fermion propagator,
  they must be included in the expression for the squared $W-$fusion graphs
  and will therefore be a part of our final result. (Actually their numerical
  significance in the final result is negligible.)
 Similar results are also obtained for the third term of Eq.~(\ref{3rd}) with
 the replacement $k_1 \leftrightarrow k_2$.
 
\bigskip

 Finally we collect together all terms that do not contain any 
 fermion propagators, that is, all those that resemble the
 structure of a squared  $W-$fusion graph, and omit all others.
 In this manner we are able to define a part of the squared matrix element 
  specific to $W-$fusion:
  \be 
\overline{|T|}^2 = \overline{|T|}^2_{ww} + \ldots \; .
 \ee
This squared amplitude will play the r\^ole of the signal for
  $W-$fusion, while all remaining squares and interferences will be
 identified as background. We stress again that this
separation between signal and background
  is now meaningful since both parts 
  are gauge invariant and well behaved at high energies.   
 The signal squared amplitude for $W-$fusion  is explicitly given by
\bea
\overline{|T|}^2_{ww}&= &\frac{1}{4}\left[ 
T^{F\mu\nu}T^{\ast F}_{\mu\nu} -M_Z^2 A_1\cdot A_1^{\ast}
-M_Z^2 A_2\cdot A_2^{\ast} +(1/4-8c_w^2)|F_1|^2 \right.
\nonumber \\
&& + |F_2|^2 \left(|A|^2-2 M_Z^4/|D_H|^2-4 M_W^2O/|D_H|^2 \right)\nonumber \\
&&\left.+ \Re e (F_1F_2^{\ast}A^{\ast})  
-2 M_Z^2\Re e \left( hF_2^{\ast}/D_H^{\ast}  \right) 
\right]       \; ,
\label{result}
\eea
where $h=h_1+h_2$ and 
\be
O= (q^2-M_H^2)(W_1+W_2)(1/D_1+1/D_2) \; .
\ee  
We observe that the $O$  term changes sign 
above and below the resonance as an 
interference term usually does. 
Finally, the unpolarized $W-$fusion cross section can be calculated from
\be
d\hat{\sigma}_{ww} = \frac{g^4}{2s}\overline{|T|}^2_{ww} 
\frac{1}{2} d\Phi_4
\label{crosssection}
\ee

Since in arriving at the result of 
Eq.~(\ref{result}) we have accounted for all possible
exchange of terms and cancellations between $W-$fusion and
$T_{Wf}+T_{ff}$ graphs, this expression should now respect unitarity.
By direct numerical calculation (see below)
we can see that this is indeed the case;
Eq.~(\ref{crosssection}) exhibits good high energy behaviour. In fact the
cross section 
assumes a slow logarithmic growth with increasing scattering energy,
exactly as in the production of an
{\it on-shell} Higgs. Furthermore the Higgs resonance is clearly 
 exhibited in the differential distribution 
 $d\sigma/dM_{ZZ}$ where $M_{ZZ}=(k_1+k_2)^2=q^2$. 

 Since the kinematics allow $q^2$ to be equal to $M_H^2$, 
 in order to carry out the numerical evaluation of Eq.~(\ref{crosssection})
 the propagator of the Higgs boson  must be regulated.
 We adopt here the naive approach of adding
 an imaginary part, equal to the total decay width of the Higgs,
 to all the Higgs  propagator denominators in Eq.~(\ref{result}).
 Specifically, we make the substitution
 \be
 D_H=q^2-M_H^2+i M_H \Gamma_H \; .
 \label{addGH}
 \ee
The width $\Gamma_H$ is taken to be
the sum of all partial, constant decay widths of the (Standard Model)
Higgs boson, each one contributing as the relevant threshold is crossed.
 With all fermions apart from
 the top quark taken as massless, this is given by\footnote{Recall
that we are interested only in heavy
Higgs bosons, decaying to $Z^0Z^0$ final states, whose total width is
dominated by the $t \bar t$, $W^+W^-$ and $Z^0Z^0$ contributions.}
 \be
 \Gamma_H = \theta (q^2-4m_t^2)\; \Gamma_H^{(tt)}
   + \theta(q^2-4M_W^2)\; \Gamma_H^{(WW)}
    +\theta (q^2-4M_Z^2)\; \Gamma_H^{(ZZ)} \; .
\label{conwidth}
 \ee
Including a constant width in this manner 
 is to some extent an {\it ad hoc} assumption, but we have found that it
works very well numerically.

 For a more consistent approach 
one has to rely on field theory.  
 In field theory, regulators for resonant
amplitudes are naturally provided through  resummation of the 
self-energy diagrams which form a geometric series.
The minimal approach, sufficient to regulate the resonance, is to 
include only the imaginary part of the 
one-loop  self energy in the resummation.
This corresponds to replacing the inverse Higgs propagator by
\be
D_H=q^2-M_H^2 \rightarrow q^2-M_H^2 + i \Im m\Pi_{HH}(q^2)
\label{addImPi}
\ee
thus giving rise to a running width 
\be
\Im m\Pi_{HH}(q^2)= 
\sqrt{q^2}\; \Gamma_H(q^2) \; .
\ee
Since the bosonic parts of the Higgs self energy are gauge dependent,
the pinch technique must be used in both the construction and 
 resummation of the self energies \cite{PaPi}. This results in a 
 gauge independent running
 width that coincides, at $q^2=M_H^2$,
 with the physical decay width of Eq.~(\ref{conwidth}),  term by term,
 viz. 
 \be
\Im m\Pi_{HH}^{(tt)}(M_H^2)=M_H\Gamma_H^{(tt)},~~~
\Im m\Pi_{HH}^{(WW)}(M_H^2)=M_H\Gamma_H^{(WW)},~~~
\Im m\Pi_{HH}^{(ZZ)}(M_H^2)=M_H\Gamma_H^{(ZZ)} \; .
 \ee
 The resummation of the Higgs self energy alone distorts
 the Higgs related  Ward identities such as Eq.~(\ref{WIHiggs}).
  This is because now the inverse Higgs propagator contains terms
  of ${\cal O}(\alpha)$. To compensate for this and still maintain
  the Ward identities,  we must also include in our amplitude 
  the one-loop imaginary parts of the relevant Higgs vertices.
  The full form of the Ward identity of Eq.~(\ref{WIHiggs}) is \cite{PaPi}
 \be
k_1^{\mu}k_2^{\nu}\Gamma^{HZZ}_{\mu\nu}(q,-k_1,-k_2)+
M_Z^2\Gamma^{H\chi\chi}(q,-k_1,-k_2)=
\frac{igM_Z}{2c_w}\left[
D_{H}(q^2)-D_{\chi}(k_1^2)-D_{\chi}(k_2^2)
\right]   \; ,
\label{fullWIHiggs}
\ee
where $\chi$ is the Goldstone boson related to the $Z$ gauge boson. 
A similar Ward identity holds for the $\Gamma_{\mu\nu}^{HWW}$ vertex. 
These are Ward identities relating the tree level vertices and 
propagators of the classical Lagrangian. They hold true before including 
ghost and gauge fixing terms. They are still valid at one loop for the 
pinch technique Green's functions. At tree level, setting $D_{\chi}(k^2)=k^2$
in Eq.~(\ref{fullWIHiggs}), 
since $\chi$ is massless before quantization, one recovers Eq.~(\ref{WIHiggs}).

However these Ward identities are in fact irrelevant for the unitarity
cancellations that we have considered, i.e. those
between the $W-$fusion graphs and the graphs $T_{Wf}$ or $T_{ff}$.
They {\it will} become essential when one attempts to separate out
the Higgs graph alone from
the rest of the $W-$fusion graphs,  and to require for it
good high energy behaviour.
In this case $D_H$ in Eq.~(\ref{fullWIHiggs}) will cancel the Higgs
propagator and thus extract a piece from the Higgs graph that combines 
with the gauge boson graphs, while the rest of the 
terms in Eq.~(\ref{fullWIHiggs}), namely
$\Gamma^{H\chi\chi}$ and $D_{\chi}$,
will remain in the Higgs graph. 
This is a cancellation `internal' to the
$W-$fusion graphs.
The extra  vertex terms in the Higgs graph 
will however  modify its contribution by ${\cal O}(\alpha)$. 
 It is conceivable that these 
 terms, if not correctly included, may lead again to violation of unitarity. 
 However since they are ${\cal O}(\alpha)$
 this would come about only at extremely high energies.

In our approach we have neglected all such terms of
${\cal O}(\alpha)$
in the numerator of the Higgs graph. However, in order to 
obtain an estimate of the difference between   
our naive treatment
and the correct treatment on the resonance, we have calculated
the  differential cross section using both a constant width
 and the pinch technique running width for the Higgs.
 We have only found negligible  numerical differences of 
${\cal O}(10^{-3}-10^{-4})$. These do not affect any of the plots we present
below. In general we believe that the numerical significance of these
terms is very small.

\bigskip

The squared amplitude of Eq.~(\ref{result}) is calculated with
{\tt FeynCalc} and a {\tt Fortran} output in terms of scalar products
between momenta and fermion currents is obtained. 
The phase space integration is done by Monte Carlo methods using
{\tt VEGAS}.
We decompose the phase space according to the structure of the Higgs
graph,
i.e. to the product of
 a three body phase space times a two body decay of the Higgs:
\be
d\Phi_4(q_1,q_2;q_3,q_4,k_1,k_2)=(2\pi)^3 d\Phi_3(q_1,q_2;q_3,q_4,q)\ dq^2 \ 
d\Phi_2(q;k_1,k_2) \; .
\ee

\setcounter{equation}{0}
\section{The process $pp\rightarrow Z Z + 2\; {\rm jets}\; + X$}

In this section we  investigate the quantitative impact of our
results by studying the  realistic case of heavy Higgs production at the LHC.
In particular, we focus on the $ZZ$ final state
 (the `gold-plated' decay channel
for a heavy Standard Model Higgs, see for example Ref.~\cite{BOOK})
We require, in addition, two forward `tag' jets \cite{Cahn86}, so that the
leading-order subprocess is $qq \to qqZZ$. We will compare the cross sections
obtained using the full scattering 
electroweak amplitude for this process \cite{GloBa}
with those obtained using the pinch-approximated $W-$fusion amplitude,
as defined and calculated in the previous section.

 The full cross section is obtained by summing over all possible
 parton subprocesses folded with the appropriate parton distributions: 
\be
\sigma (pp\rightarrow Z Z + 2j + X) = 
\sum\int dx_1 dx_2 f_{q_1/p}(x_1,Q^2)f_{q_2/p}(x_2,Q^2)
\hat\sigma(q_1q_2\rightarrow q'_1q'_2 Z Z) \; .
\label{fullsigma}
\ee
For the parton distributions we use the default MRST set \cite{mrs98},
with scale choice 
$Q=M_H$. 
 As our primary aim is to compare cross sections
calculated using different scattering amplitudes, 
we fix the parton and parameter
choices throughout the study, 
and impose the same cuts on the final-state particles.
Because it is  physically indistinguishable, we must 
include also the resonant $Z-$fusion contribution. 
The $Z-$fusion amplitude can be obtained
 straightforwardly from the relevant 3 Feynman graphs, since they form a 
 gauge invariant subset. 
In some of the   subprocesses one of the two resonant  mechanisms 
appears as Higgs-strahlung  $S$ graphs, which are suppressed relative 
to the fusion graphs by an additional factor of $\hat{s}$. 
 In such cases
 we retain only the energetically dominant, fusion one.
 For example, in 
 $\bar{u}u \rightarrow \bar{d}dZZ$ the 
Higgs-strahlung  $S_{ZZ}$
 graphs are suppressed.
   When they are both relevant, as in 
   $     d u \rightarrow      u dZZ$ for example, 
    we will neglect the interference between them. 
This is because the momenta of the final state quarks are crossed 
   in the $Z-$fusion graphs relative to the $W-$fusion quark momenta. 
   Since each
   amplitude peaks in the forward region there is a negligibly small
   region of overlap in phase space. 
   For the same reason, when calculating amplitudes with identical fermions 
   in the final state we will not consider the crossed diagrams, including 
    only the direct one without the symmetry factor $1/2$. 

We begin by showing, in Fig.~\ref{fig7}, the total subprocess cross
section as a function of the subprocess centre-of-mass energy
$\sqrt{s}$, calculated three different ways.
 For purposes of illustration we take $M_H = 500$~GeV.
The solid line (F) correponds to the six $W-$ fusion graphs
calculated in the Feynman gauge. As discussed in the Introduction,
this exhibits the unitarity-violating behaviour $\sigma \sim s^2$
at high energy. Isolating the Higgs resonance graph alone (dotted line
H) again leads to unitarity-violating behaviour, but now
 $\sigma \sim s$.\footnote{Notice also the threshold behaviour of
this contribution at $\sqrt{s} \simeq M_H$.}
Finally, applying the pinch technique as described in the previous
section yields
acceptable high-energy behaviour of the form $\sigma \sim \ell n(s)$
(dash-dotted line PT).

The `bad' high-energy behaviour of the $W-$fusion graphs completely
swamps the Higgs resonance behaviour. This is illustrated in
Fig.~\ref{fig8}, which shows the $ZZ$ invariant mass distribution
at two values of the subprocess energy $\sqrt{s}$, again for
$M_H = 500$~GeV. For the $W-$fusion graphs contribution,
we see the resonance at $500$~GeV
disappear under the background as $\sqrt{s}$
increases. At high $\sqrt{s}$ the cross section is spread approximately
uniformly over the range $(0,\sqrt{s})$.
In contrast, applying the pinch technique gives
invariant mass distributions in which the resonance is clearly visible at both
values of $\sqrt{s}$ considered.
Of course the extent to which the Higgs resonance is visible over
the underlying non-resonant diagram contributions depends on $\Gamma_H$.
Figure~\ref{fig9} shows the same $ZZ$
(pinch-technique) mass distribution as in Fig.~\ref{fig8}, for $M_H =
500$~GeV and $\sqrt{s} = 4$~TeV, together with the corresponding
distributions for $M_H = 750$~GeV and $1000$~GeV. Notice that,
as expected, the resonant peak broadens as $M_H$ increases, becoming
barely visible over the non-resonant contributions at $M_H = 1000$~GeV.

\bigskip

Our final two figures show distributions for the {\it full}
($\sqrt{s} = 14$ TeV)
proton-proton cross section, i.e. with parton distributions
folded in as in Eq.~(\ref{fullsigma}).
In these figures we compare the pinch-technique
(approximate) result with the full all-diagrams calculation of
Ref.~\cite{GloBa}. Figure~\ref{fig10} shows the $ZZ$ invariant mass
distributions for $M_H=500$~GeV and $M_H = 740$~GeV.
We see that the resonance region is indeed very well approximated by the
pinch-technique result. The high-mass tail is also in good agreement
with the full result. Only the low $M_{ZZ}$ region below ${\cal O}(400
\; {\rm GeV})$ shows any significant difference. Here the large number
of non-$W-$fusion graphs in the full calculation leads to an excess
over the pinch-technique result. We note, however, that this low-mass
region would in all likelihood be removed by experimental cuts in a
realistic analysis.

It is relevant to ask whether other features of the final state are
well approximated by the pinch-technique method. As an illustration,
we consider the transverse momentum distribution of the forward jets
accompanying the $Z$ boson pair. Since it may be possible to detect
these jets in the LHC experiments, it is important that the
pinch-technique result gives an accurate description of their
production properties. Figure~\ref{fig11} shows the predictions for the
inclusive jet $p_T$ distribution in the full calculation and the
pinch-technique approximation, for $M_H = 500$ and $740$~GeV.
A lower $M_{ZZ}$ cut is imposed (see above). The distributions are
indeed very similar, particularly in shape. (The small
pinch-technique excess
can be traced back to the slightly higher $M_{ZZ}$ distribution
in Fig.~\ref{fig10}.) The rapidity distributions of the jets (not shown)
are also very similar. This gives us confidence that all important kinematic
features of the $ZZ+ 2$~jets production process are well reproduced
by the pinch-technique method.

\section{Conclusions}

In this paper we have used the pinch-technique to define and calculate a 
$W$-fusion subamplitude of the full amplitude for the electroweak process 
$qq\rightarrow qqZZ$. The pinch-technique amplitude is based on the 
$W$-fusion subset of diagrams, and includes the Higgs resonance contribution, 
i.e. $WW\rightarrow H \rightarrow ZZ$.  
More importantly, the pinch-technique amplitude squared is 
{\it gauge invariant } and the corresponding cross section exhibits 
{\it good high energy behaviour}. In addition, our numerical studies 
show that the pinch-technique gives an excellent approximation to the 
full calculation for such quantities as the $ZZ$ invariant mass distribution, 
particularly in the region of the resonance, and the rapidity and transverse 
momentum distributions of the jets accompanying the $ZZ$ pair. 
Not surprisingly, the expression for the pinch-technique is much more 
compact than that for the full amplitude, with a corresponding computation 
time per event  which is between one and two orders of magnitude faster. 

In summary, we have demonstrated, in an non-trivial example, that 
a {\it meaningful} and {\it well defined} 
separation between {\it signal} and {\it background} can be achieved even when 
they contribute {\it coherently} to the same final state. In the particular 
case of interest the {\it signal} turns out to be dominant 
while its numerical computation compared to the full cross section has been 
expedited enormously.  
We believe, therefore, that the pinch-technique amplitude could be a useful 
simulation and analysis tool for Higgs production via 
$W$-fusion at the LHC.

\vskip 1truecm

\noindent{\bf Acknowledgements}
\medskip

\noindent 
We would like to thank 
 Joannis Papavassiliou for useful discussions and Nigel Glover 
for providing us with his {\tt Fortran} code of the full calculation. 
K.P. would also like to thank Matthias Heyssler for help with the 
drawing of  
Feynman graphs  and Marco Stratmann for technical support and 
help with {\tt PAW}. 
This work was supported in part by the EU Programme
``Human Capital and Mobility'', Network 
``Phenomenological Studies of Electroweak and Strong Interactions
at Future Colliders'', contract CHRX-CT93-0319 (DG 12 COMA). The work of 
K.P. is  also supported by a {\it Marie Curie} TMR grant FMBI-CT96-1033. 
 
\goodbreak


\begin{figure}[ht]
\begin{center}
\mbox{\epsfig{figure=
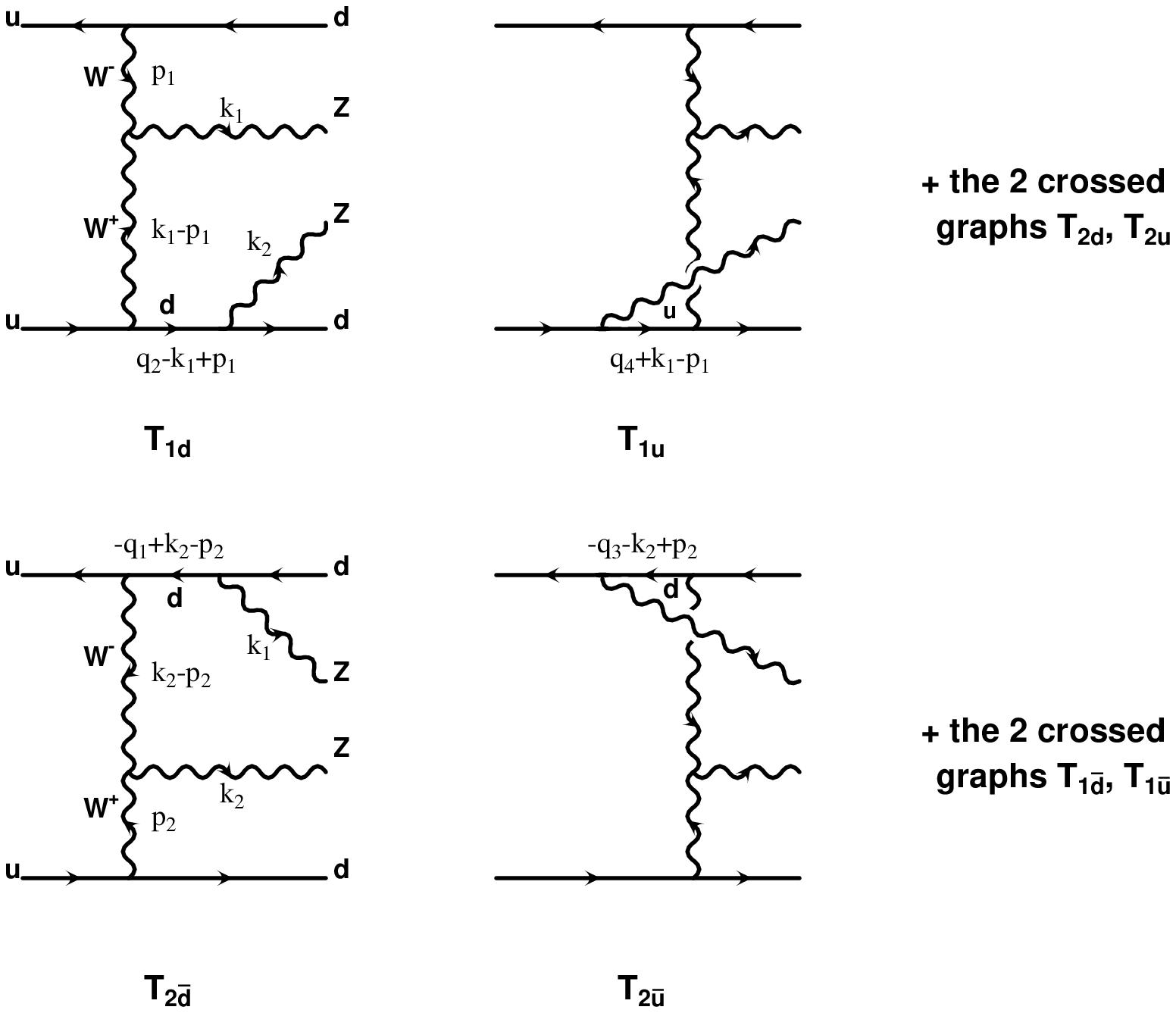,width=15.0cm}}
\label{fig2}
\end{center}
\vspace*{1cm}
\noindent {\bf Figure 2:} The $T_{Wf}$ graphs. The relevant
kinematics are also shown. Crossed graphs are omitted.
\end{figure}

\begin{figure}[ht]
\begin{center}
\mbox{\epsfig{figure=
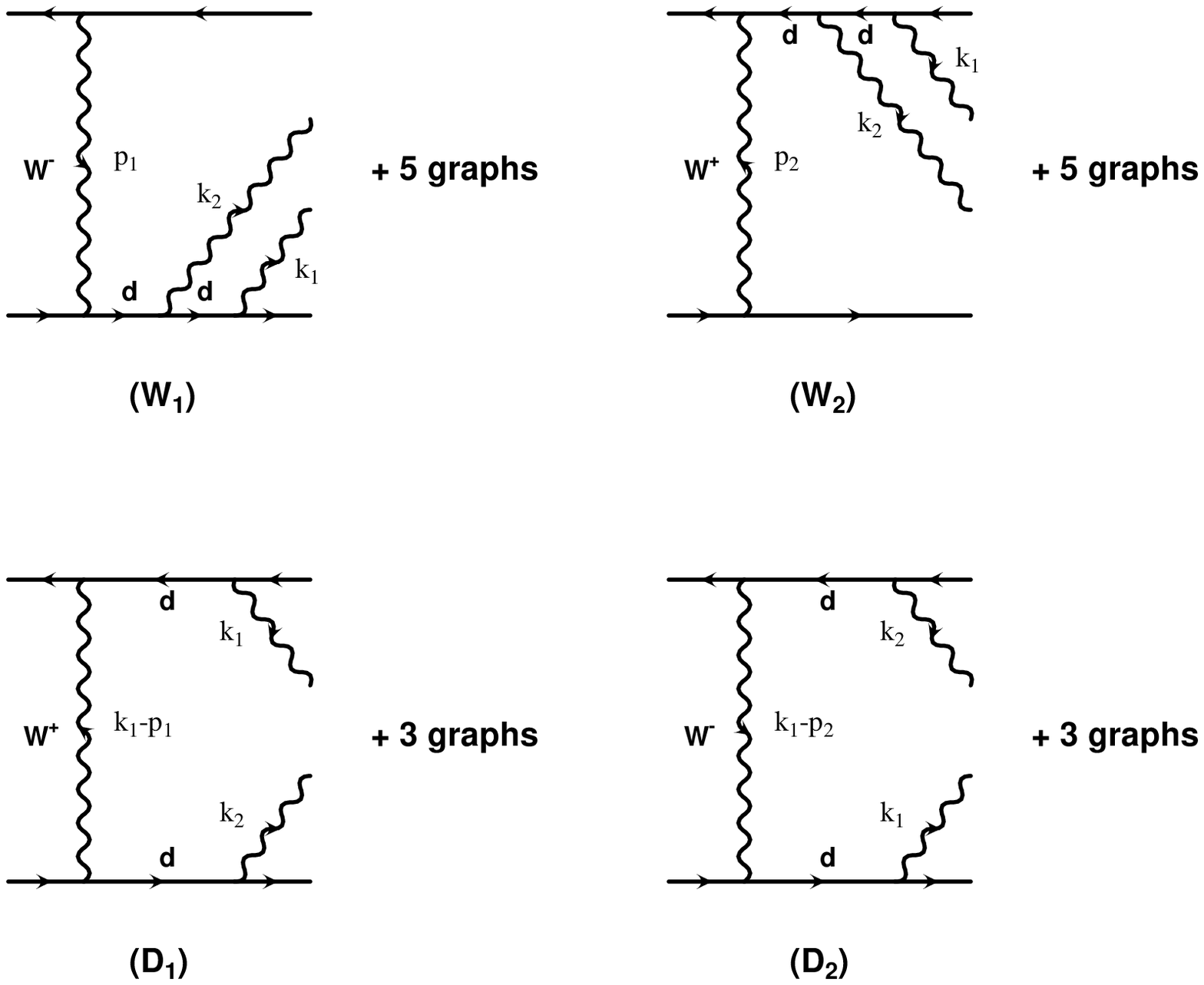,width=15.0cm}}
\label{fig3}
\end{center}
\noindent {\bf Figure 3:} The $T_{ff}$ graphs. The graphs not shown refer
to permutations of the relative positions of the $Z$ bosons
on the fermion line to which they are attached.
\vspace*{2cm}
\end{figure}

\begin{figure}[ht]
\begin{center}
\mbox{\epsfig{figure=
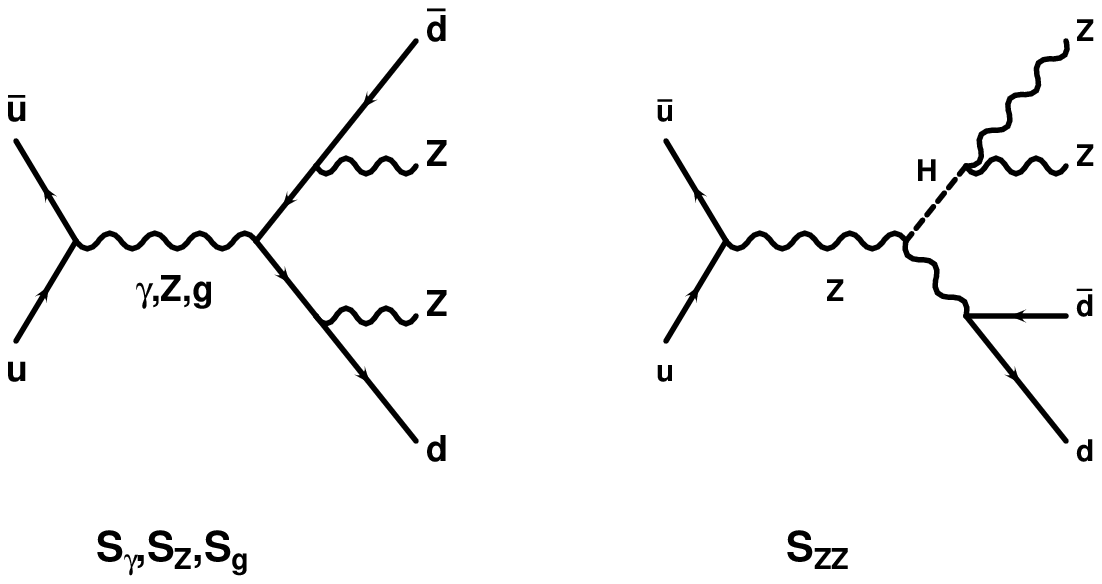,width=15.0cm}}
\label{fig4}
\end{center}
\noindent {\bf Figure 4:} The $S$ graphs. There are 20 $S_{g}$,
 $S_{\gamma}$ and $S_{Z}$  graphs respectively. One of the
three Higgs-strahlung graphs  is also shown.
\vspace*{2cm}
\end{figure}

\begin{figure}[ht]
\begin{center}
\mbox{\epsfig{figure=
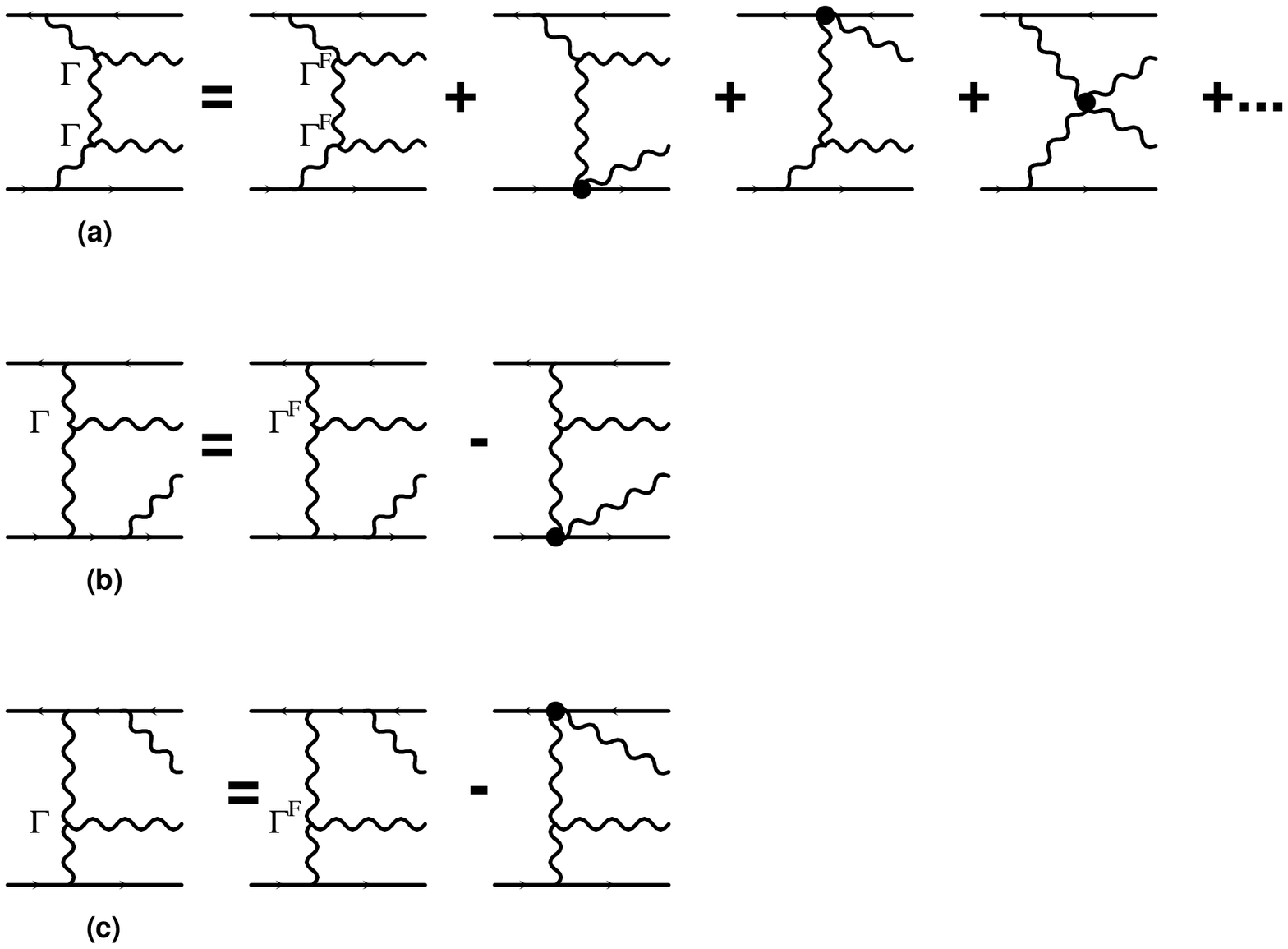,width=15.0cm}}
\label{fig5}
\end{center}
\noindent {\bf Figure 5:} Pinch terms induced by the presence of the
trilinear gauge vertices.
\vspace*{2cm}
\end{figure}

\begin{figure}[ht]
\begin{center}
\mbox{\epsfig{figure=
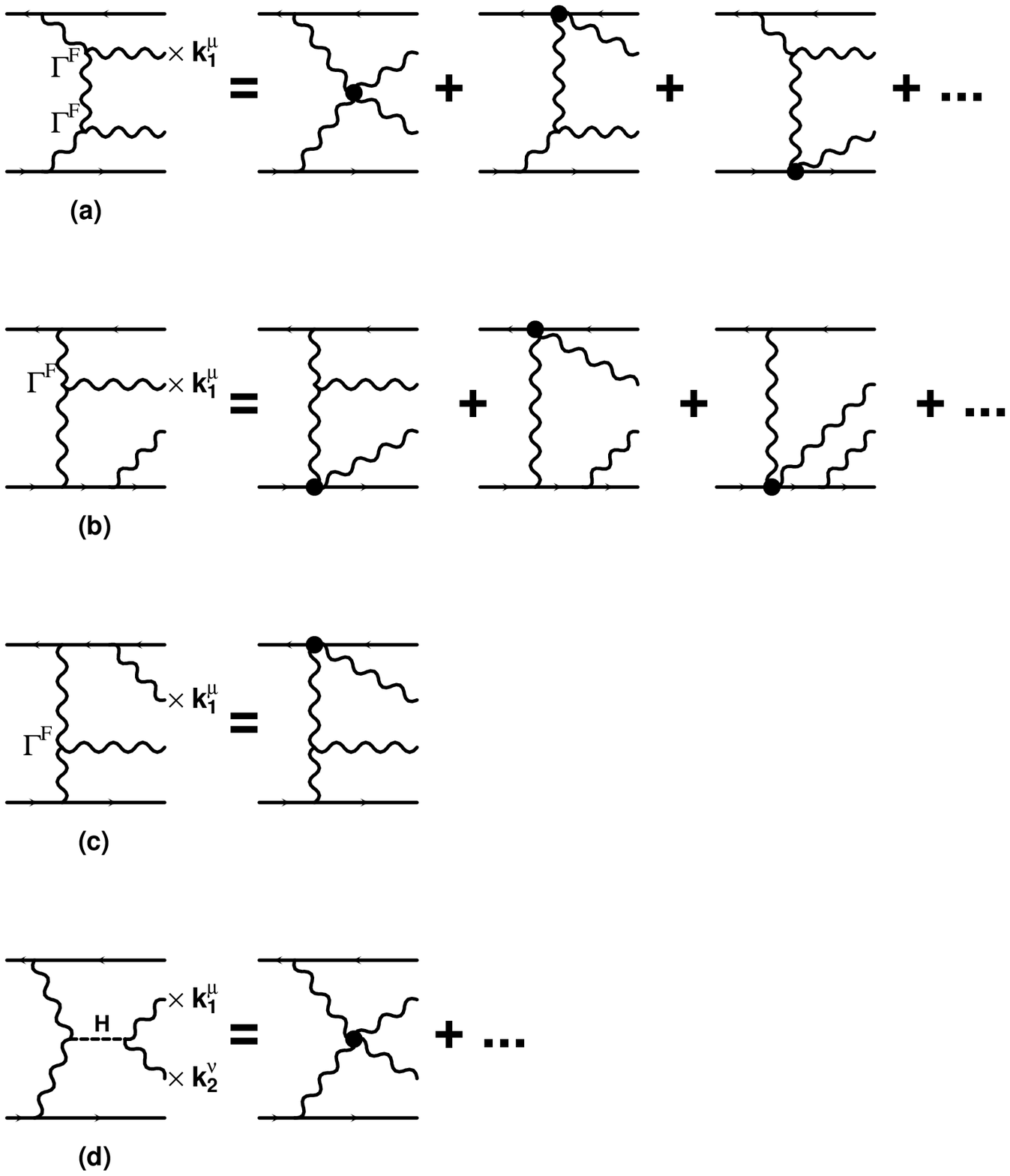,width=15.0cm}}
\label{fig6}
\end{center}
\noindent {\bf Figure 6:} Pinch terms induced by the momentum of a longitudinal 
$Z$ boson. The sum of $(a)$, $(b)$, $(c)$ and the relevant term for 
the $T_{ff}$ graphs is a pictorial representation of the Ward Identity of 
Eq.~(\ref{wiz}) when the trilinear vertices in the graphs are the full ones.
In the sum, all pinch terms on the r.h.s. cancel and the remainder equals
the graphs ${\cal T}(\chi)$, with the external $Z$ replaced by its
would-be Goldstone boson $\chi$.
Terms represented by the ellipsis on the r.h.s.
of $(b)$ contribute to the remainder $R_1^{\nu}$ of
Eq.~(\ref{k1T}). 
In $(d)$ we show how pinching is induced in the Higgs graph according to the  
Ward identity of Eq.~(\ref{WIHiggs}). The pinch term on the r.h.s. contributes
to the $g_{\alpha\beta}$ term of  $F_1$  in 
Eq.~(\ref{k2k1T}).
\vspace*{2cm}
\end{figure}

\begin{figure}[ht]
\begin{center}
\mbox{\epsfig{figure=
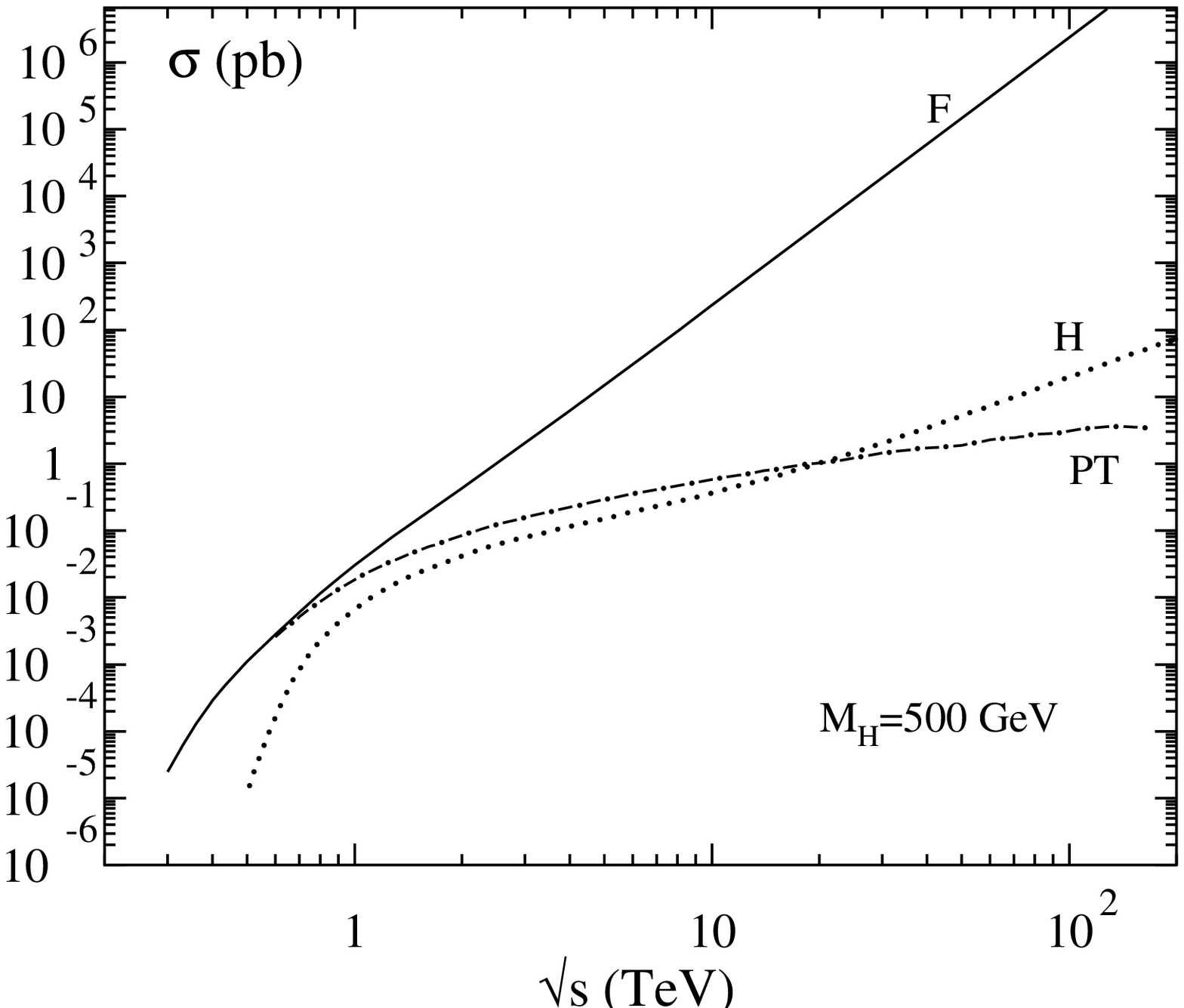,width=17.5cm}}
\label{fig7}
\end{center}
\vspace*{1cm}
\noindent {\bf Figure 7:}
Total subprocess cross section for $W-$fusion. The solid line represents the
six  $W-$fusion graphs in the Feynman gauge.
The dotted line is the Higgs graph alone. The dash-dotted line is
the pinch technique result.
\vspace*{2cm}
\end{figure}

\begin{figure}[ht]
\begin{center}
\mbox{\epsfig{figure=
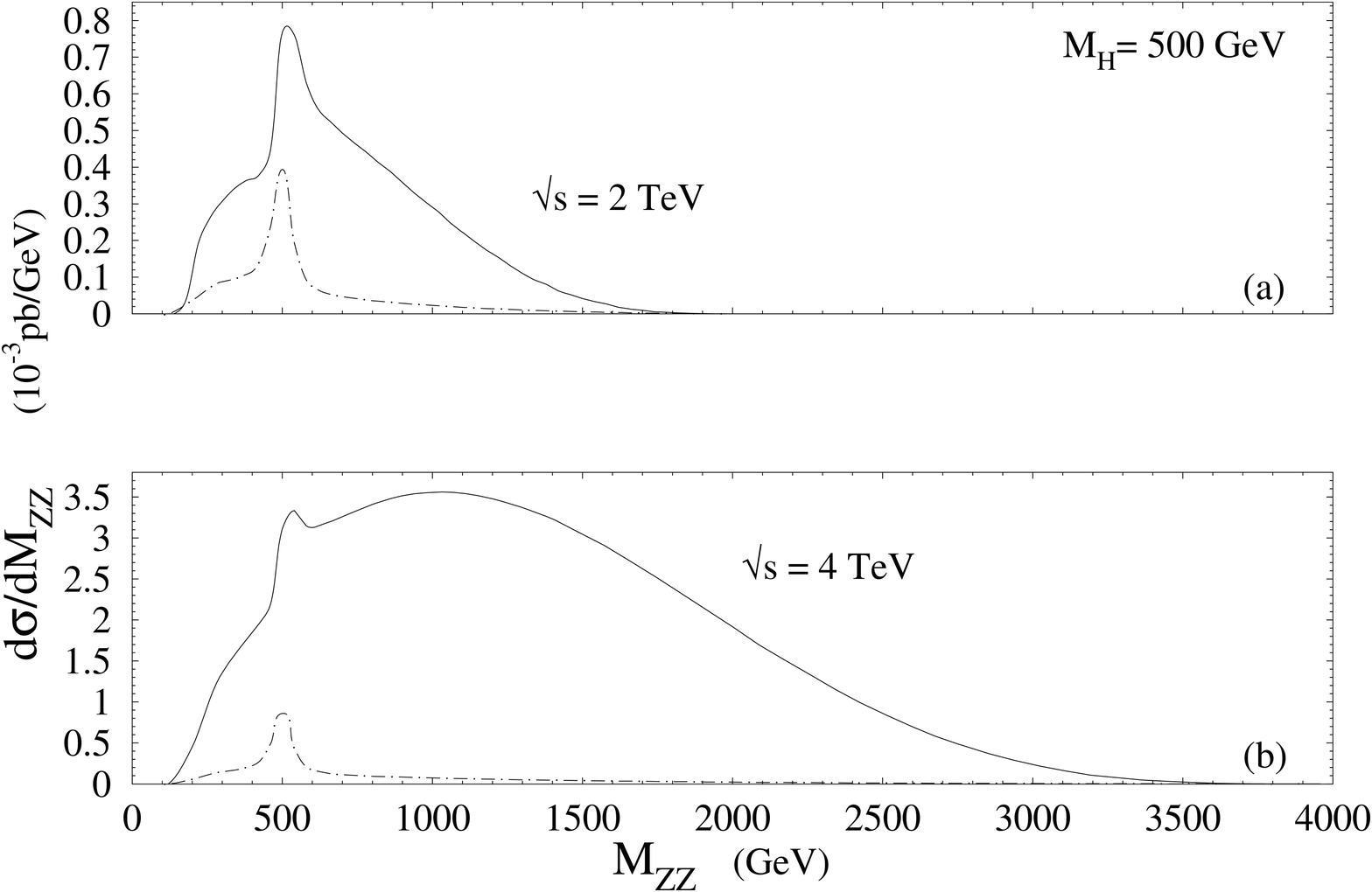,width=17.5cm}}
\label{fig8}
\end{center}
\vspace*{1cm}
\noindent {\bf Figure 8:} Invariant mass distribution of the $Z$ boson pair.
The solid line is the pinch-technique result while the dotted line is the
result of the 6 $W-$fusion graphs.
\vspace*{2cm}
\end{figure}

\begin{figure}[ht]
\begin{center}
\mbox{\epsfig{figure=
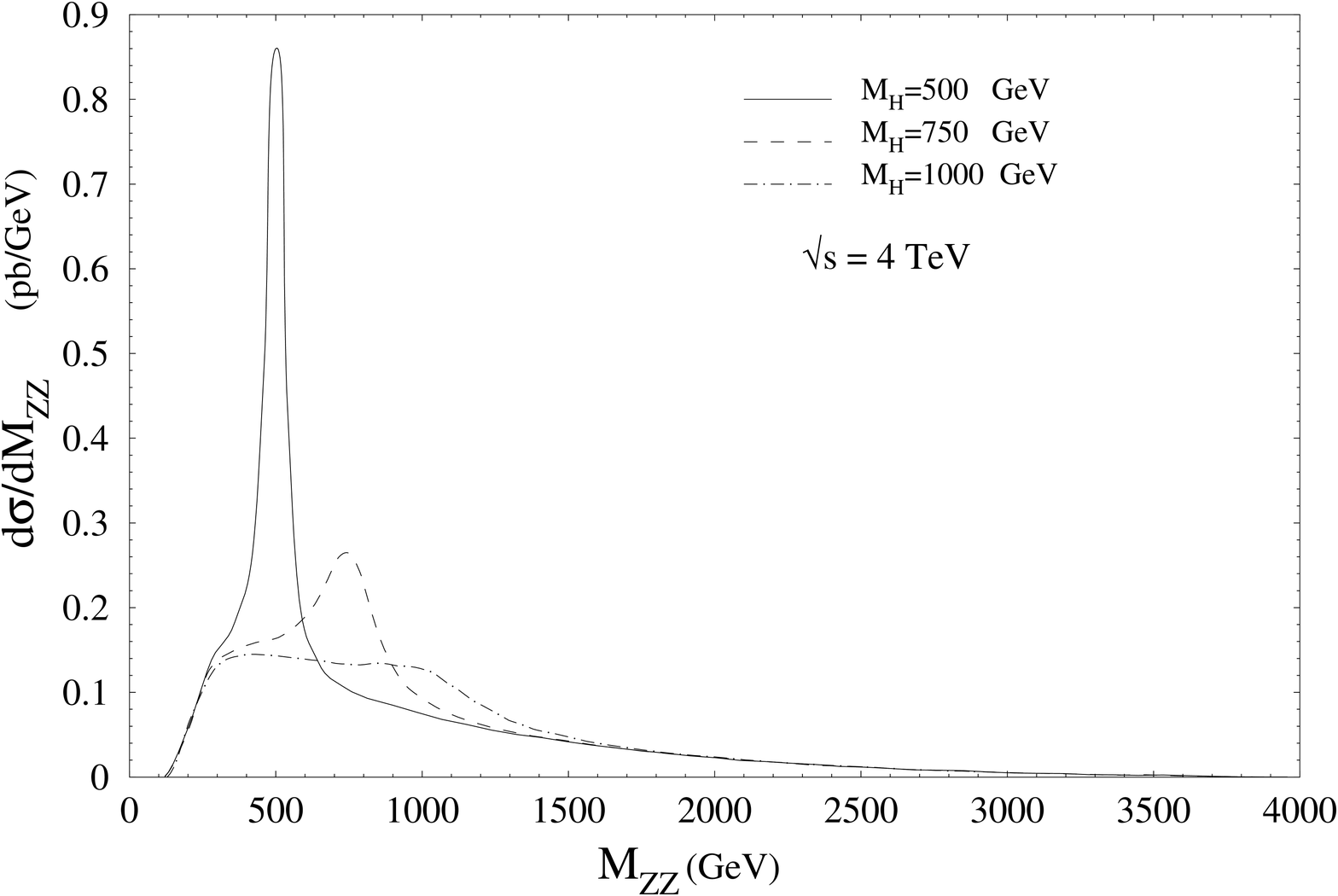,width=17.5cm}}
\label{fig9}
\end{center}
\vspace*{1cm}
\noindent {\bf Figure 9:} 
Invariant mass distribution for the $Z$ boson pair of 
the pinch-technique result, for different values of the Higgs mass
and for subprocess scattering energy $\sqrt{s} = 4$~TeV.
\vspace*{2cm}
\end{figure}

\begin{figure}[ht]
\begin{center}
\mbox{\epsfig{figure=
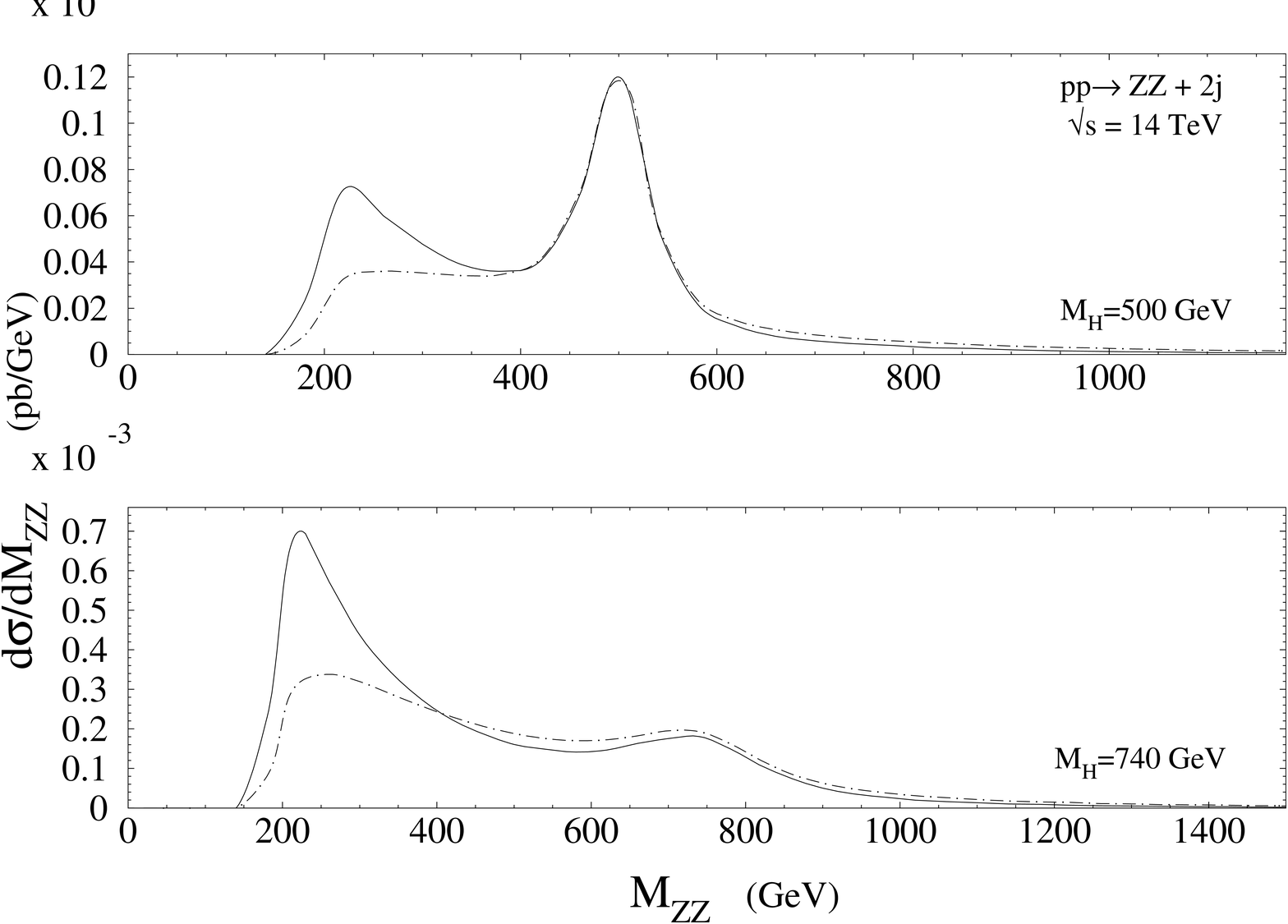,width=17.5cm}}
\label{fig10}
\end{center}
\vspace*{1cm}
\noindent {\bf Figure 10:} 
Invariant mass distribution of the $Z$ boson pair for 
$pp\rightarrow Z Z + 2\; {\rm jets}\; + X$ at the LHC.
The solid line represents
the full calculation in which all electroweak graphs are included.
The dashed-dotted line is the pinch-technique approximation
in which only the resonant $W-$fusion and $Z-$fusion graphs are retained.
\vspace*{2cm}
\end{figure}

\begin{figure}[ht]
\begin{center}
\mbox{\epsfig{figure=
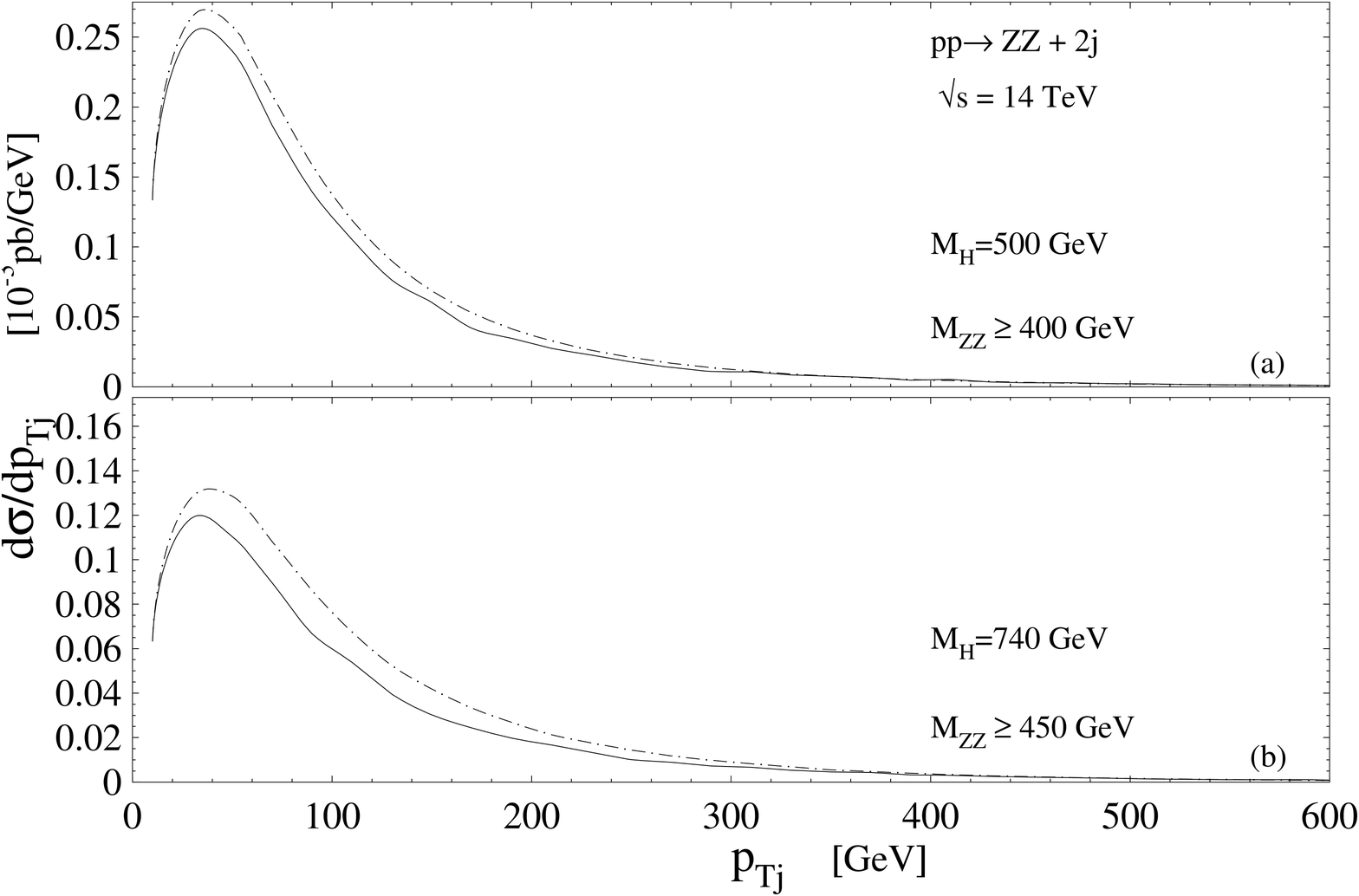,width=17.5cm}}
\label{fig11}
\end{center}
\vspace*{1cm}
\noindent {\bf Figure 11:} 
Distributions of the jet transverse momentum for
$pp\rightarrow Z Z + 2\; {\rm jets}\; + X$ at the LHC.
The solid line represents 
the full calculation in which all electroweak graphs are included.
The dashed-dotted line is our approximation 
in which only the resonant $W-$fusion and $Z-$fusion graphs are retained.
\vspace*{2cm}
\end{figure}

\end{document}